\documentclass[twocolumn]{aastex631}

\usepackage{amsmath,bm}

\DeclareMathOperator\arctanh{arctanh}

\newcommand{\beq}{\begin{equation}}           
\newcommand{\eeq}{\end{equation}}             
\newcommand{\bfi}{\begin{figure}}           
\newcommand{\efi}{\end{figure}}             

\shortauthors{Karamazov, Timko, \& Heyrovsk\'y}

\begin{document}

\title{Gravitational Lensing By a Massive Object in a Dark Matter Halo. I. Critical Curves and Caustics}
	
\author[0000-0002-7919-499X]{Michal Karamazov}
\affiliation{Institute of Theoretical Physics, Faculty of Mathematics and Physics, Charles University,\\ V Hole\v{s}ovi\v{c}k\'ach 2, 18000~Praha 8, Czech Republic}
\author{Luk\'a\v{s} Timko}
\affiliation{Institute of Theoretical Physics, Faculty of Mathematics and Physics, Charles University,\\ V Hole\v{s}ovi\v{c}k\'ach 2, 18000~Praha 8, Czech Republic}
\author[0000-0002-5198-5343]{David Heyrovsk\'y}
\affiliation{Institute of Theoretical Physics, Faculty of Mathematics and Physics, Charles University,\\ V Hole\v{s}ovi\v{c}k\'ach 2, 18000~Praha 8, Czech Republic}
\email{michal.karamazov@gmail.com,\; timkolukas@seznam.cz}
\email{david.heyrovsky@mff.cuni.cz}

\begin{abstract}

We study the gravitational lensing properties of a massive object in a dark matter halo, concentrating on the critical curves and caustics of the combined lens. We model the system in the simplest approximation by a point mass embedded in a spherical Navarro--Frenk--White density profile. The low number of parameters of such a model permits a systematic exploration of its parameter space. We present galleries of critical curves and caustics for different masses and positions of the point in the halo. We demonstrate the existence of a critical mass, above which the gravitational influence of the centrally positioned point is strong enough to eliminate the radial critical curve and caustic of the halo. In the point-mass parameter space we identify the boundaries at which critical-curve transitions and corresponding caustic metamorphoses occur. The number of transitions as a function of position of the point is surprisingly high, ranging from three for higher masses to as many as eight for lower masses. On the caustics we identify the occurrence of six different types of caustic metamorphoses. We illustrate the peculiar properties of the single radial critical curve and caustic appearing in an additional unusual non-local metamorphosis for a critical mass positioned at the halo center. Although we constructed the model primarily to study the lensing influence of individual galaxies in a galaxy cluster, it can also be used to study the lensing by dwarf satellite galaxies in the halo of a host galaxy, as well as (super)massive black holes at a general position in a galactic halo.

\end{abstract}

\section{INTRODUCTION}
\label{sec:Intro}

Galaxy clusters provide a unique setting in which gravitational lensing plays an important role, uncovering information about the background universe as well as about the cluster itself \citep{kneib_natarajan11}. On the one hand, clusters act as gravitational telescopes for studying the more distant populations of high-redshift galaxies and protogalaxies. On the other hand, combined analyses of weak and strong gravitational lensing can be used to map the total mass distribution in the cluster \citep[e.g.,][]{finney_etal18,jauzac_etal18}. A number of different techniques have been developed for joining the statistical information on weak image deformations in the outer parts of the cluster with the information on specific multiply imaged systems in its inner parts \citep{meneghetti_etal17}.

These techniques have advanced to a state in which properties of analyzed clusters can be compared with properties of simulated clusters formed in cosmological structure-formation simulations. In a recent study, \cite{meneghetti_etal20} compared the lensing effects of substructure in a set of observed clusters and in their simulated counterparts. They found a substantial discrepancy in the population of small-scale gravitational lenses: their lensing efficiency in the observed clusters was more than an order of magnitude higher than in the simulated clusters. Lacking an obvious single explanation for this surprising result, \cite{meneghetti_etal20} suggested its possible resolution might involve either systematic issues with simulations, or incorrect assumptions about dark matter properties.

Instead of simulating lensing by an advanced realistic model of a galaxy cluster with all its different components, in this work we take a first step in a bottom-up approach. We study the lensing effect of a single massive object in the dark matter halo of a galaxy cluster. Using a simple model with few parameters allows us to systematically explore the lensing behavior of the system and its parameter-space variations. Results of such a study can be used as a stepping stone to exploring the properties of more advanced models. At the same time, they may aid the interpretation of the local lensing behavior in the vicinity of individual galaxies in a cluster.

We model the mass distribution of the cluster by a spherical Navarro--Frenk--White (hereafter NFW) density profile \citep{navarro_etal96}, which has been shown to describe adequately the combined dark matter and baryonic gas distribution in galaxy clusters \citep{newman_etal13}. For our purposes, the NFW profile has the additional advantage of yielding a simple analytic expression for the gravitational deflection angle. This in turn permits an analytic derivation of the Jacobian and other lensing quantities, as well as more efficient inverse-ray-shooting computations. We model the massive object using the simplest approximation, i.e., that of a point mass. While this is a rather poor model for describing a galaxy, at a sufficient distance the gravitational field of any massive object can be described by its monopole. At the scale of the galaxy cluster it is not unreasonable as a first approximation.

The results of structure-formation simulations indicate that the NFW profile is suitable not just for galaxy-cluster halos, but more generally for dark matter halos down to the scale of individual galaxies \citep{ludlow_etal13}. In view of this finding, our lens model can be used just as well for studying two other astrophysical scenarios. First, it can describe the lensing effect of a substructure or a dwarf satellite galaxy in the dark matter halo of a host galaxy \citep[e.g.,][]{hezaveh_etal16}. Second, it can describe the lensing by a (super)massive black hole in a galaxy \citep[e.g.,][]{mao_etal01,bowman_etal04}. We note that observational data on galactic dark-matter distributions indicate a preference for profiles with a central core rather than a cusp \citep{salucci19}. Nevertheless, the NFW profile can be used as a reasonable approximation for the halos of elliptical \citep[e.g.,][]{shajib_etal21} or even massive spiral galaxies \citep[e.g.,][]{rodrigues_etal17}.

In this article we describe the basic lensing properties of the model, focusing on the structure of its critical curves and caustics. In a companion article we explore the effect of the point mass on the shear and on the images formed by the lens. The content presented here is organized as follows. In Section~\ref{sec:NFW} we give a brief overview of lensing by a NFW halo, illustrating the dependence of critical-curve and caustic radii on the halo convergence parameter. We select a fiducial value of the parameter and use it for computing most of the subsequently presented results. In Section~\ref{sec:NFWP} we explore the properties of the combined NFW halo + point-mass lens model. For a centrally positioned point mass we study the different lensing regimes as a function of its mass in Section~\ref{sec:central_position}. For a general position of the point mass we explore the critical-curve transitions and caustic metamorphoses and map the corresponding boundaries in the parameter space of the point mass in Section~\ref{sec:general_position}. In Section~\ref{sec:discussion} we comment on the effect of varying the halo convergence parameter, and discuss the relevance of the results in different astrophysical scenarios. We summarize our main findings in Section~\ref{sec:summary}. In Appendix~\ref{sec:Appendix-analytic} we present useful analytic results and approximations. In Appendix~\ref{sec:Appendix-vanishing_curves} we describe the unusual lensing properties for a critical-mass point positioned at the halo center.

\section{LENSING BY A NAVARRO--FRENK--WHITE HALO}
\label{sec:NFW}

\subsection{Density profile and convergence}
\label{sec:NFW-density}

The three-dimensional density profile of a spherical dark-matter halo can be described by the Navarro--Frenk--White profile \citep{navarro_etal96},
\beq
\rho(r)=\rho_{\text{s}}\,\left(\frac{r}{r_{\text{s}}}\right)^{\mkern-10mu -1}\left(1+\frac{r}{r_{\text{s}}}\right)^{\mkern-10mu -2}\,,
\label{eq:NFW_density}
\eeq
where $r$ is the three-dimensional radial distance from the center, $r_{\text{s}}$ is the scale radius and $\rho_{\text{s}}$ is a characteristic density such that $\rho(r_{\text{s}})=\rho_{\text{s}}/4$. At small radii $r\ll r_{\text{s}}$ the density diverges as $\rho\sim r^{-1}$, at large radii $r\gg r_{\text{s}}$ the density drops as $\rho\sim r^{-3}$, and at the scale radius ${\rm d}\ln{\rho}\,/{\rm d}\ln{r}\,|_{r=r_{\text{s}}}=-2$. The halo mass enclosed in a sphere of radius $r$ is
\begin{multline}
M(r)=4\pi\int_0^r\,r'^{\,2}\rho(r')\,{\rm d}\;\!r'\\= 4\pi r_{\text{s}}^3 \rho_{\text{s}} \left[\ln{\left(1+\frac{r}{r_{\text{s}}}\right)}-\frac{r}{r+r_{\text{s}}} \right]\,.
\label{eq:NFW_mass}
\end{multline}
Due to the logarithmic divergence of the NFW halo mass for $r\gg r_{\text{s}}\,$, the profile is extended only to a certain distance such as $r_{200}$, at which the mean density within the enclosed sphere is 200 times the critical density at the redshift of the halo, $\rho_{\text{crit}}(z)$. The ratio of the two characteristic radii defines the concentration parameter of the halo, $c_{\text{s}}=r_{200}/r_{\text{s}}$.

Despite its density divergence at the center and its mass divergence at large radii, for a broad range of intermediate radii the NFW profile presents a good fit to cold dark matter halo profiles. For galaxy cluster halos this agreement has been demonstrated by cluster lensing analyses \citep{okabe_etal13,umetsu_diemer17} or by X-ray emission analyses \citep{ettori_etal13}.

In order to compute light deflection by the NFW halo we first integrate Equation~(\ref{eq:NFW_density}) along the line of sight to obtain the NFW surface density. We express $r=r_{\text{s}} \sqrt{x^2+l^2}$ in terms of the distances $x$ projected in the plane of the sky and $l$ along the line of sight, both in units of the scale radius $r_{\text{s}}$. The convergence
\beq
\kappa(x)=\frac{r_{\text{s}}}{\Sigma_{\text{cr}}}\int_{-\infty}^{\infty}\,\rho(r_{\text{s}} \sqrt{x^2+l^2})\, {\rm d}\;\!l
\label{eq:kappa_integral}
\eeq
is defined as the surface density expressed in units of the critical surface density
\beq
\Sigma_{\text{cr}}=\frac{c^2}{4\pi G}\frac{D_{\text{s}}}{D_{\text{l}}\,D_{\text{ls}}}\,,
\label{eq:critical_density}
\eeq
where $c$ is the speed of light, $G$ the gravitational constant, and $D_{\text{l}},\, D_{\text{s}}$, and $D_{\text{ls}}$ are the angular diameter distances from the observer to the lens (i.e., the halo in our case), from the observer to the source of light (e.g., a background galaxy or quasar), and from the lens to the source, respectively.

For the NFW density from Equation~(\ref{eq:NFW_density}) the integral in Equation~(\ref{eq:kappa_integral}) can be performed analytically \citep[e.g.,][]{bartelmann96,wright_brainerd00,keeton01,golse_kneib02}, yielding the NFW convergence as a function of the plane-of-the-sky radial position $x$,
\beq
\kappa(x)=2\,\kappa_{\text{s}}\;\frac{1-\mathcal{F}(x)}{x^2-1}\,,
\label{eq:NFW_kappa}
\eeq
where the dimensionless halo convergence parameter $\kappa_{\text{s}}= \rho_{\text{s}} \, r_{\text{s}} /\Sigma_{\text{cr}}$ and the function
\beq
\mathcal{F}(x)=\begin{cases}
\frac{\displaystyle\arctanh{\sqrt{1-x^2}}}{\displaystyle\sqrt{1-x^2}} & \text{for $x<1$}\,,\\
\hfil 1 & \text{for $x=1$}\,,\\
\frac{\displaystyle\arctan{\sqrt{x^2-1}}}{\displaystyle\sqrt{x^2-1}} & \text{for $x>1$}\,.
\end{cases}
\label{eq:f(x)}
\eeq
The NFW convergence decreases monotonically and smoothly with radial position $x$ throughout its range. As shown in Equation~(\ref{eq:NFW_kappa_origin}), it has a logarithmic divergence for $x \to 0$, where $\kappa(x)\approx -2\,\kappa_{\text{s}}\,\ln{x}$. It drops to $\kappa(1)=2\,\kappa_{\text{s}}/3$ at the scale radius, as shown in Equation~(\ref{eq:NFW_kappa_radius}), and decreases further to zero as $\kappa(x)\approx 2\,\kappa_{\text{s}} \,x^{-2}$ for $x\gg 1$. The unit-convergence radius $x_0$ has a special significance from the perspective of lensing. It can be determined for a given value of $\kappa_{\text{s}}$ by setting $\kappa(x_0)=1$ in Equation~(\ref{eq:NFW_kappa}),
\beq
\frac{1-x_0^2}{\mathcal{F}(x_0)-1}=2\,\kappa_{\text{s}}\,,
\label{eq:x0_radius}
\eeq
which can be solved numerically. The solutions are illustrated and discussed further in Section~\ref{sec:NFW-Jacobian}.

\subsection{Lens equation}
\label{sec:NFW-lens_equation}

The gravitational field of the halo deflects a light ray passing at point $\boldsymbol x$ in the plane of the sky by the deflection angle
\beq
\boldsymbol \alpha(\boldsymbol x)=\frac{4\,G\, M_{\text{cyl}}(x\,r_{\text{s}})}{c^2\,r_{\text{s}}}\;\frac{\boldsymbol x}{x^2}\,,
\label{eq:NFW_angle}
\eeq
where $M_{\text{cyl}}(x\,r_{\text{s}})$ is the mass within a radius $x\,r_{\text{s}}$ along the line of sight through the halo center,
\begin{multline}
M_{\text{cyl}}(x\,r_{\text{s}})=2\,\pi\,r_{\text{s}}^2\,\Sigma_{\text{cr}}\int_{0}^{x}\,\kappa(x')\,x'\,{\rm d}\;\!x'\\=\, 4\,\pi\,r_{\text{s}}^3\,\rho_{\text{s}}\,\left[\ln{\frac{x}{2}}+\mathcal{F}(x)\right]\,,
\label{eq:NFW_mass_cylinder}
\end{multline}
where we used the NFW convergence from Equation~(\ref{eq:NFW_kappa}). The deflection angle for the NFW halo thus is
\begin{multline}
\boldsymbol \alpha(\boldsymbol x)=\frac{16\,\pi\,G\,r_{\text{s}}^2\,\rho_{\text{s}}}{c^2}\,\left[\ln{\frac{x}{2}}+\mathcal{F}(x)\right]\,\frac{\boldsymbol x}{x^2}\\=\,\frac{4\,\kappa_{\text{s}}\,r_{\text{s}}\,D_{\text{s}}}{D_{\text{l}}\,D_{\text{ls}}}\,\left[\ln{\frac{x}{2}}+\mathcal{F}(x)\right]\,\frac{\boldsymbol x}{x^2}\,,
\label{eq:NFW_deflection}
\end{multline}
where we used Equation~(\ref{eq:critical_density}) and the definition of $\kappa_{\text{s}}$ under Equation~(\ref{eq:NFW_kappa}) to get the second expression. The deflection angle can be used in the general lens equation,
\beq
\boldsymbol \beta = \boldsymbol \theta - \frac{D_{\text{ls}}}{D_{\text{s}}}\boldsymbol \alpha\,,
\label{eq:lens_equation}
\eeq
connecting the angular position $\boldsymbol \beta$ of a background source with the angular position $\boldsymbol \theta$ of its image formed by the lens. In gravitational lensing the angles are often expressed in units of the Einstein radius\footnote{Here given as the radius of the tangential critical curve of the mass $M_{\text{NFW}}$ as a point lens. For a lens with a NFW density profile the radius of the tangential critical curve has to be computed numerically from Equation~(\ref{eq:tangential}).} of the lens,
\beq
\theta_{\text{E}}=\sqrt{\frac{4\,G\,M_{\text{NFW}}}{c^2}\,\frac{D_{\text{ls}}}{D_{\text{l}}\,D_{\text{s}}}}\,,
\label{eq:Einstein_radius}
\eeq
where we set $M_{\text{NFW}}=M(c_{\text{s}}\,r_{\text{s}})$ as the total mass within radius $r_{200}$ using Equation~(\ref{eq:NFW_mass}),
\beq
M_{\text{NFW}}= 4\,\pi\,r_{\text{s}}^3\,\rho_{\text{s}}\,[\,\ln{(1+c_{\text{s}})}-c_{\text{s}}/(1+c_{\text{s}})\,]\,.
\label{eq:NFW_mass_total}
\eeq
If we introduce the angular scale radius in units of the Einstein radius $\theta_{\text{s}}^*=r_{\text{s}}/(D_{\text{l}}\,\theta_{\text{E}})$, we may write the lens equation for the NFW profile as
\begin{multline}
\boldsymbol \beta^*=\boldsymbol \theta^*-[\,\ln{(1+c_{\text{s}})}-c_{\text{s}}/(1+c_{\text{s}})\,]^{-1}\\ \times \left[\ln{\frac{\theta^*}{2\,\theta_{\text{s}}^*}}+\mathcal{F}\left(\frac{\theta^*}{\theta_{\text{s}}^*}\right)\right]
\,\frac{\boldsymbol \theta^*}{(\theta^*)^2}\,,
\label{eq:NFW_lens_equation_Einstein}
\end{multline}
where $\boldsymbol \beta^*=\boldsymbol \beta / \theta_{\text{E}}=$ and $\boldsymbol \theta^* =\boldsymbol \theta / \theta_{\text{E}}$. Alternatively, we may express the angles in units of the angular scale length of the halo, $r_{\text{s}}/D_{\text{l}}$. In these units the source position $\boldsymbol y = \boldsymbol \beta\, D_{\text{l}}/r_{\text{s}}$ and the image position $\boldsymbol x = \boldsymbol \theta\, D_{\text{l}}/r_{\text{s}}$, so that the lens equation has the form
\beq
\boldsymbol y=\boldsymbol x - 4\,\kappa_{\text{s}}\,
\left[\ln{\frac{x}{2}}+\mathcal{F}(x)\right]\,\frac{\boldsymbol x}{x^2}\,.
\label{eq:NFW_lens_equation}
\eeq
Equation~(\ref{eq:NFW_lens_equation_Einstein}), which is expressed in familiar lensing units, explicitly involves two parameters, $c_{\text{s}}$ and $\theta_{\text{s}}^*$. In the rest of this work we use the more compact Equation~(\ref{eq:NFW_lens_equation}), which involves a single parameter $\kappa_{\text{s}}$ that is related to the two parameters by
\beq
\kappa_{\text{s}} = (2\,\theta_{\text{s}}^*)^{-2}\,[\,\ln{(1+c_{\text{s}})}-c_{\text{s}}/(1+c_{\text{s}})\,]^{-1}\,.
\label{eq:kappa_s_conversion}
\eeq
This expression can be derived using Equations~(\ref{eq:critical_density}), (\ref{eq:Einstein_radius}), and (\ref{eq:NFW_mass_total}) in the definition of $\kappa_{\text{s}}$ above Equation~(\ref{eq:f(x)}).

We illustrate the conversion in Figure~\ref{fig:clusters} by the dotted $\kappa_{\text{s}} = \mathrm{const.}$ contours in a $\theta_{\text{s}}^*$ vs. $c_{\text{s}}$ plot including sample observational data. The plot range is set to include the parameter combinations of 19 observed galaxy clusters from the CLASH survey \citep{merten_etal15}, marked here by the crosses. We use the Merten et al. cluster scale radii $r_{\text{s}}$, masses $M_{200c}$ for $M_{\text{NFW}}$, concentrations $c_{200c}$ for $c_{\text{s}}$, and cluster redshifts $z$ to compute the angular diameter distance $d_{\text{A}}(z)=D_{\text{l}}$ in a FLRW universe with PLANCK 2015 cosmological parameters \citep{planck_etal15}. To obtain $\theta_{\text{s}}^*$ we compute the angular Einstein radius $\theta_{\text{E}}$ for asymptotically distant sources, replacing $D_{\text{ls}}/D_{\text{s}}\rightarrow 1$. This replacement does not hold exactly for angular diameter distances, for which the asymptotic ratio is lens-redshift dependent. However, the approximation overestimates the Einstein radii of the clusters in the sample merely by $3\text{--}11\,\%$.

\subsection{Jacobian, critical curve, caustic}
\label{sec:NFW-Jacobian}

Many important lensing quantities are obtained by computing the Jacobian of the lens equation: the inverse of its absolute value yields the magnification of a point-source image at $\boldsymbol x$, its sign indicates the image parity, and its zero contour defines the critical curve, which in turn yields the caustic when mapped back to the source-plane positions $\boldsymbol y$.

For the NFW profile, the Jacobian of Equation~(\ref{eq:NFW_lens_equation}) is
\begin{multline}
\mathrm{det}\,J(\boldsymbol x)=\left| \frac{\partial\,\boldsymbol y}{\partial\,\boldsymbol x} \right| = \left\{ 1-\frac{4\,\kappa_{\text{s}}}{x^2}\, \left[\ln{\frac{x}{2}}+\mathcal{F}(x)\right] \right\}\\ \times \left\{ 1+\frac{4\,\kappa_{\text{s}}}{x^2}\, \left[\ln{\frac{x}{2}}+\mathcal{F}(x)\right] -4\,\kappa_{\text{s}}\,\frac{\mathcal{F}(x)-1}{1-x^2} \right\}\,,
\label{eq:NFW_Jacobian}
\end{multline}
which, due to symmetry, is a purely radial function of image position $x$. For $x\to 0$ at the halo center $\mathrm{det}\,J(\boldsymbol x)\approx4\,\kappa_{\text{s}}^2\,\ln^2{x}\to\infty$, as shown in Equation~(\ref{eq:NFW_Jacobian_origin}). For $x\to\infty$ the Jacobian asymptotically reaches unity, $\mathrm{det}\,J(\boldsymbol x)\to 1$.

The factorized form of the Jacobian indicates that the critical curve $\mathrm{det}\,J(\boldsymbol x)=0$ consists of solutions of two simpler equations. The factor in the first braces yields the tangential critical curve, which is a circle $|\boldsymbol x|=x_{\text{T}}$ with radius obtained by numerically solving
\beq
1-\frac{4\,\kappa_{\text{s}}}{x_{\text{T}}^2}\, \left[\,\ln{\frac{x_{\text{T}}}{2}}+\mathcal{F}(x_{\text{T}})\,\right]=0\,,
\label{eq:tangential}
\eeq
an equation with a single solution for any value of $\kappa_{\text{s}}$. If we introduce the mean convergence within a circle of radius $x$,
\beq
\bar{\kappa}(x)=\frac{1}{\pi\,x^2}\,\int_0^x 2\pi\,x'\kappa(x')\,{\rm d}\;\!x'=\frac{4\,\kappa_{\text{s}}}{x^2}\, \left[\,\ln{\frac{x}{2}}+\mathcal{F}(x)\,\right]\,,
\label{eq:mean_convergence}
\eeq
where we took into account Equation~(\ref{eq:NFW_mass_cylinder}), we see that Equation~(\ref{eq:tangential}) implies $\bar{\kappa}(x_{\text{T}})=1$. The tangential critical curve thus encloses a circle with unit mean convergence. Recalling that the NFW convergence is a monotonically decreasing function, this means that $\kappa(x_{\text{T}})<\bar{\kappa}(x_{\text{T}})=1$ and, thus, $x_{\text{T}} > x_0$, where the radius of unit convergence $x_0$ is given by Equation~(\ref{eq:x0_radius}). Substituting the critical curve $\boldsymbol x=x_{\text{T}}\,(\cos{\varphi},\sin{\varphi})$ with $\varphi\in [\,0, 2\pi\,]$ in Equation~(\ref{eq:NFW_lens_equation}) and using Equation~(\ref{eq:tangential}), we obtain the corresponding part of the caustic:
\beq
\boldsymbol y = \boldsymbol x\,\left\{1-\frac{4\,\kappa_{\text{s}}}{x_{\text{T}}^2}\, \left[\,\ln{\frac{x_{\text{T}}}{2}}+\mathcal{F}(x_{\text{T}})\,\right]\right\}=(\,0,0\,)\,.
\label{eq:tangential_caustic}
\eeq
The tangential part of the caustic consists of a single point at the origin.

Setting the second braces in Equation~(\ref{eq:NFW_Jacobian}) equal to zero yields the radial critical curve. This is another circle with radius $x_{\text{R}}$ obtained by numerically solving
\beq
1+\frac{4\,\kappa_{\text{s}}}{x_{\text{R}}^2}\, \left[\,\ln{\frac{x_{\text{R}}}{2}}+\mathcal{F}(x_{\text{R}})\,\right]-4\,\kappa_{\text{s}}\,\frac{\mathcal{F}(x_{\text{R}})-1}{1-x_{\text{R}}^2}=0\,,
\label{eq:radial}
\eeq
which also has a single solution for any $\kappa_{\text{s}}$. The l.h.s. can be written as a simple combination of Equation~(\ref{eq:mean_convergence}) and Equation~(\ref{eq:NFW_kappa}), yielding
\beq
1+\bar{\kappa}(x_{\text{R}})-2\,\kappa(x_{\text{R}})=0\,.
\label{eq:radial_rewritten}
\eeq
Therefore, the convergence at the radius of the radial critical curve
\beq
\kappa(x_{\text{R}})=\frac{1}{2}\,\left[\,1+\bar{\kappa}(x_{\text{R}})\,\right]
\label{eq:radial_convergence}
\eeq
is the average of the mean convergence at the radius and 1. For the monotonically decreasing NFW convergence $\bar{\kappa}(x_{\text{R}})>\kappa(x_{\text{R}})>1$ and, thus, $x_{\text{R}} < x_0$. Substituting the radial critical curve $\boldsymbol x=x_{\text{R}}\,(\cos{\varphi},\sin{\varphi})$ with $\varphi\in [\,0, 2\pi\,]$ in Equation~(\ref{eq:NFW_lens_equation}) and using Equation~(\ref{eq:radial}), we obtain the corresponding part of the caustic,
\begin{multline}
\boldsymbol y = \boldsymbol x\,\left\{1-\frac{4\,\kappa_{\text{s}}}{x_{\text{R}}^2}\, \left[\,\ln{\frac{x_{\text{R}}}{2}}+\mathcal{F}(x_{\text{R}})\,\right]\right\}\\ =-2\,\left[\, 2\,\kappa_{\text{s}}\,\frac{\mathcal{F}(x_{\text{R}})-1}{1-x_{\text{R}}^2}-1\,\right]\,x_{\text{R}}\,(\cos{\varphi},\sin{\varphi})\,,
\label{eq:radial_caustic}
\end{multline}
where the expression in the square brackets is equal to $\kappa(x_{\text{R}})-1$, which is positive. The radial part of the caustic is thus a circle with radius
\beq
y_{\text{R}}=2\,x_{\text{R}}\,[\,\kappa(x_{\text{R}})-1\,]\,.
\label{eq:yR}
\eeq

\bfi
{\centering
\vspace{0cm}
\hspace{0cm}
\includegraphics[width=8.5 cm]{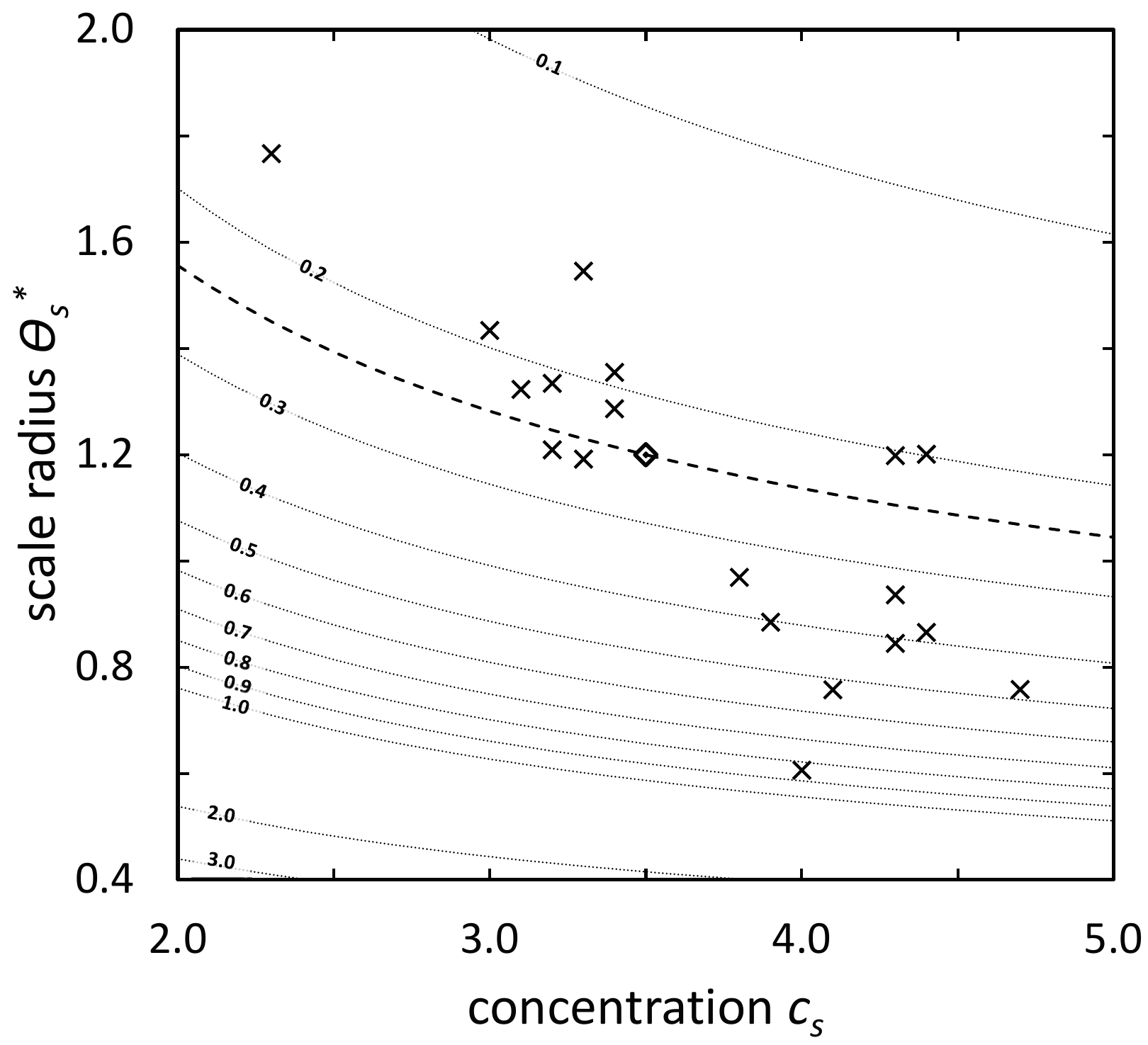}
\caption{Contours of Navarro--Frenk--White halo convergence parameter $\kappa_{\text{s}}$ as a function of concentration parameter $c_{\text{s}}$ and scale radius in units of Einstein radius $\theta_{\text{s}}^*$, expressed by Equation~(\ref{eq:kappa_s_conversion}). Crosses indicate galaxy cluster data from the \cite{merten_etal15} sample. The diamond marks the parameter combination $\{c_{\text{s}},\,\theta_{\text{s}}^*\}=\{3.5,\,1.2\}$; the bold dashed contour passing through it corresponds to the value $\kappa_{\text{s}}\approx 0.239035$ used for a fiducial NFW halo in this work.}
\label{fig:clusters}}
\efi

\bfi
{\centering
\vspace{0cm}
\hspace{0cm}
\includegraphics[width=8.5 cm]{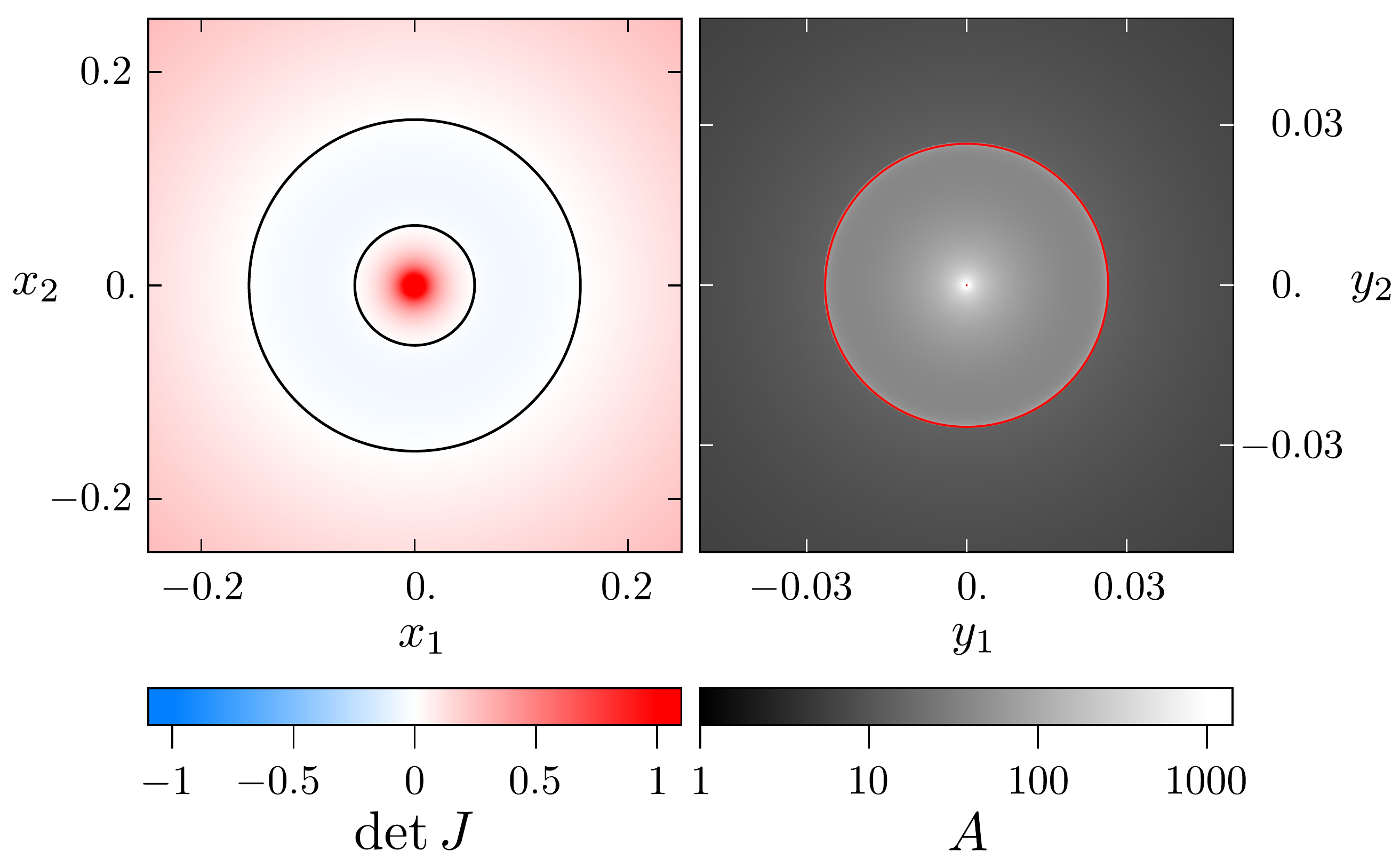}
\caption{Critical curve (left panel, black) and caustic (right panel, red) of a NFW halo with $\kappa_{\text{s}}\approx 0.239035$. The larger tangential critical curve corresponds to the central point-like caustic; the smaller radial critical curve corresponds to the circular caustic. The image-plane color map in the left panel shows the lens-equation Jacobian $\mathrm{det}\,J(\boldsymbol x)$, with colors saturating at $|\mathrm{det}\,J|=1$ (the Jacobian is divergent at the origin). The source-plane grayscale map in the right panel shows the total point-source magnification $A(\boldsymbol y)$, with white saturating at $A=1000$ (the magnification is divergent at the caustic).}
\label{fig:NFW}}
\efi

Figure~\ref{fig:NFW} shows the critical curve in the image plane (black, left panel) and caustic in the source plane (red, right panel) for a NFW density profile with $\kappa_{\text{s}}\approx 0.239035$. This value was obtained from Equation~(\ref{eq:kappa_s_conversion}) for the combination
\beq
\{c_{\text{s}},\,\theta_{\text{s}}^*\}=\{3.5,\,1.2\}
\label{eq:parameter_choice}
\eeq
chosen to represent the \cite{merten_etal15} cluster data and marked by the diamond in Figure~\ref{fig:clusters}. The outer tangential and inner radial critical curves are plotted over a color map of the Jacobian $\mathrm{det}\,J(\boldsymbol x)$, marked red where positive and blue where negative. Lighter areas indicate regions with higher image magnification (brighter images), darker areas regions with lower image magnification (dimmer images). Inside the radial critical curve the Jacobian is positive, increasing divergently toward the origin. Any images appearing close to the origin are thus strongly demagnified. In the annulus between the radial and tangential critical curves the Jacobian is negative, thus any images appearing here have negative parity. Outside the tangential critical curve the Jacobian is positive, increasing monotonically outward, and approaching 1 asymptotically.

The caustic is plotted over a grayscale magnification map, showing the total point-source magnification
\beq
A(\boldsymbol y)=\sum_{i} |\mathrm{det}\,J(\boldsymbol x_i)|^{-1}
\label{eq:magnification}
\eeq
of all images $\boldsymbol x_i$ formed by the lens for a point source at $\boldsymbol y$. The magnification map is computed by inverse ray shooting \citep[e.g.,][]{kayser_etal86}, with the color ranging from black for lowest magnification $A=1$ to white for highest magnification $A\geq 1000$. The magnification diverges at the caustic, here more prominently at the central point (tangential caustic).  At the circular radial caustic the magnification remains finite from the outer side but diverges from the inner side in an extremely narrow high-magnification region \citep[see also][]{martel_shapiro03}. Clearly, sources positioned outside the radial caustic are magnified substantially less than those positioned inside.

Although Figure~\ref{fig:NFW} is plotted for a single value of the NFW convergence parameter $\kappa_{\text{s}}$, the general character of the critical curve, caustic, Jacobian and magnification maps does not change for other values. What changes are the radii of the tangential and radial critical curves, $x_{\text{T}}$ and $x_{\text{R}}$, respectively, and the radial caustic radius $y_{\text{R}}$ \citep{bartelmann96,martel_shapiro03}. Figure~\ref{fig:NFW_radii} shows the dependence of these radii and the unit convergence radius $x_0$ on $\kappa_{\text{s}}$. All the plotted radii are simple monotonically increasing functions of $\kappa_{\text{s}}$. The vertical dot-dashed line indicates the value $\kappa_{\text{s}}\approx 0.239035$ chosen for illustration in Figure~\ref{fig:NFW} as well as in the rest of this work.

\bfi
{\centering
\vspace{0cm}
\hspace{-0.5cm}
\includegraphics[width=8 cm]{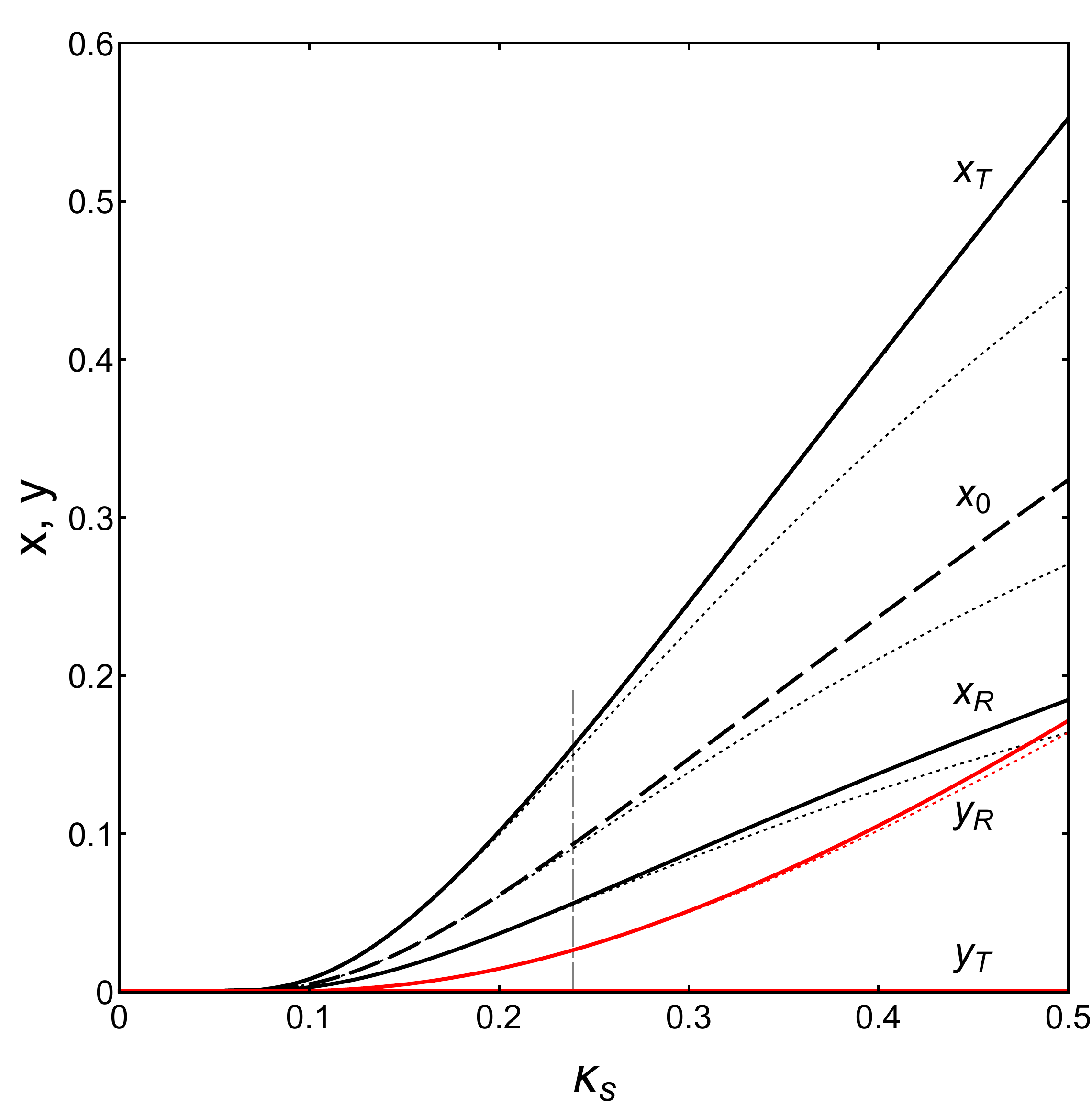}
\caption{Radii of critical curves (solid black: tangential $x_{\text{T}}$, radial $x_{\text{R}}$), caustics (solid red: tangential $y_{\text{T}}=0$, radial $y_{\text{R}}$), and the unit-convergence circle $x_0$ (dashed black) of NFW halos as a function of the convergence parameter $\kappa_{\text{s}}$. The dotted curves show the low-convergence analytic approximations given by Equation~(\ref{eq:approx_x}) and Equation~(\ref{eq:approx_y}). The dot-dashed vertical line indicates the fiducial value $\kappa_{\text{s}}\approx 0.239035$ used in Figure~\ref{fig:NFW} and the rest of this work.}
\label{fig:NFW_radii}}
\efi

While the values of the radii for general $\kappa_{\text{s}}$ plotted in Figure~\ref{fig:NFW_radii} have to be computed numerically, for $\kappa\lesssim 0.2$ they may be approximated by analytic expressions using Equation~(\ref{eq:f(x)_origin}) and Equation~(\ref{eq:ln_f(x)_origin}) from Appendix~\ref{sec:Appendix-origin} in Equation~(\ref{eq:tangential}), Equation~(\ref{eq:x0_radius}), Equation~(\ref{eq:radial}), and Equation~(\ref{eq:radial_caustic}). For the critical-curve radii \citep{dumet-montoya_etal13} and the unit-convergence radius we find
\beq
\left\{x_{\text{T}},\, x_0,\, x_{\text{R}}\right\}\approx 2\,\mathrm{e}^{-(1+3\,\kappa_{\text{s}})/(2\,\kappa_{\text{s}})}\,\left\{\mathrm{e,\, \sqrt{e},}\, 1\right\}
\label{eq:approx_x}
\eeq
and for the (radial) caustic radius we find
\beq
y_{\text{R}}\approx4\,\kappa_{\text{s}}\,\mathrm{e}^{-(1+3\,\kappa_{\text{s}})/(2\,\kappa_{\text{s}})}\,.
\label{eq:approx_y}
\eeq
These approximations are marked by dotted curves in Figure~\ref{fig:NFW_radii}. All four radii shrink exponentially fast for low values of $\kappa_{\text{s}}$. Nevertheless, the ratios between the radii $x_{\text{T}},\, x_0,\, x_{\text{R}}$ are constant in this regime, given by the factors in the braces on the r.h.s. of Equation~(\ref{eq:approx_x}).

\section{LENSING BY A NFW HALO + POINT MASS}
\label{sec:NFWP}

\subsection{Lens equation}
\label{sec:NFWP-lens_equation}

The lensing effect of an additional compact object with mass distributed in a region much smaller than the halo scale radius $r_{\text{s}}$ can be modelled by adding a point-mass deflection term
\beq
\bm{\alpha_{\text{P}}} (\boldsymbol \theta)=\frac{4\,G\, M_{\text{P}}}{c^2\,D_{\text{l}}}\;\frac{\boldsymbol \theta - \bm{\theta_{\text{P}}}}{|\boldsymbol \theta - \bm{\theta_{\text{P}}}|^2}
\label{eq:point_mass_deflection}
\eeq
to the deflection angle in Equation~(\ref{eq:lens_equation}). Here $M_{\text{P}}$ is the mass of the object and $\bm{\theta_{\text{P}}}$ its angular position from the halo center. If the object acted as an isolated lens, its region of influence could be measured by its angular Einstein radius
\beq
\theta_{\text{EP}}=\sqrt{\frac{4\,G\,M_{\text{P}}}{c^2}\,\frac{D_{\text{ls}}}{D_{\text{l}}\,D_{\text{s}}}}\,.
\label{eq:P_Einstein_radius}
\eeq
In our case, with the object embedded in the NFW halo, we will use $\theta_{\text{EP}}$ for comparison and illustration purposes.

Adding the point-mass term in units of the NFW-halo Einstein radius to Equation~(\ref{eq:NFW_lens_equation_Einstein}), we obtain the full lens equation
\begin{multline}
\boldsymbol \beta^*=\boldsymbol \theta^*-[\,\ln{(1+c_{\text{s}})}-c_{\text{s}}/(1+c_{\text{s}})\,]^{-1}\\ \times \left[\ln{\frac{\theta^*}{2\,\theta_{\text{s}}^*}}+\mathcal{F}\left(\frac{\theta^*}{\theta_{\text{s}}^*}\right)\right]
\,\frac{\boldsymbol \theta^*}{(\theta^*)^2}-\frac{M_{\text{P}}}{M_{\text{NFW}}}\,\frac{\boldsymbol \theta^* - \bm{\theta_{\text{P}}}^*}{|\boldsymbol \theta^* - \bm{\theta_{\text{P}}}^*|^2}\,,
\label{eq:NFWP_lens_equation_Einstein}
\end{multline}
where $\bm{\theta_{\text{P}}}^* = \bm{\theta_{\text{P}}} / \theta_{\text{E}}$. The influence of the added object is proportional to its relative mass and drops inversely proportionally to the angular separation from its position. Similarly, we may add the term in units of the angular scale length of the halo to Equation~(\ref{eq:NFW_lens_equation}) and obtain the lens equation in the form
\beq
\boldsymbol y=\boldsymbol x - 4\,\kappa_{\text{s}}\,
\left[\ln{\frac{x}{2}}+\mathcal{F}(x)\right]\,\frac{\boldsymbol x}{x^2}-\kappa_{\text{P}}\,\frac{\boldsymbol x-\bm{x_{\text{P}}}}{|\boldsymbol x - \bm{x_{\text{P}}}|^2}\,,
\label{eq:NFWP_lens_equation}
\eeq
where the point-mass position $\bm{x_{\text{P}}} = \bm{\theta_{\text{P}}}\,D_{\text{l}}/r_{\text{s}}$. The newly introduced dimensionless mass parameter $\kappa_{\text{P}}$ can be expressed in several equivalent ways,
\beq
\kappa_{\text{P}}=\frac{M_{\text{P}}}{M_{\text{NFW}}}\,(\theta_{\text{s}}^*)^{-2}=\frac{\theta_{\text{EP}}^2\,D_{\text{l}}^2}{r_{\text{s}}^2}= \frac{M_{\text{P}}}{\pi\,r_{\text{s}}^2\,\Sigma_{\text{cr}}}\,,
\label{eq:kappa_P}
\eeq
where we used the definition of $\theta_{\text{s}}^*$ above Equation~(\ref{eq:NFW_lens_equation_Einstein}), Equation~(\ref{eq:P_Einstein_radius}), and Equation~(\ref{eq:critical_density}). The first expression defines the transformation from the parameters appearing in Equation~(\ref{eq:NFWP_lens_equation_Einstein}). The second expression identifies $\kappa_{\text{P}}$ as the ratio of the areas (or solid angles) of the point-mass Einstein circle and the halo scale-radius circle. The third expression shows that we may interpret $\kappa_{\text{P}}$ as the convergence corresponding to the surface density of the mass $M_{\text{P}}$ spread out over the area of the halo scale-radius circle.

In the rest of this work we will use the lens equation in the more compact form provided by Equation~(\ref{eq:NFWP_lens_equation}). The equation involves four parameters: the convergences $\kappa_{\text{s}}$ and $\kappa_{\text{P}}$, and the two components of the point-mass position $\bm{x_{\text{P}}}$. For exploring the lens properties of the model we may always rotate our coordinate axes to position the point mass along the positive horizontal axis, so that $\bm{x_{\text{P}}}=(x_{\text{P}},0)$ with $x_{\text{P}}\geq 0$. With this choice of orientation only three free parameters remain: one describing the NFW halo ($\kappa_{\text{s}}$) and two describing the point mass ($\kappa_{\text{P}}$ and $x_{\text{P}}$).

\subsection{Jacobian}
\label{sec:NFWP-Jacobian}

Computing the determinant of the Jacobi matrix consisting of the partial derivatives $\partial y_i/\partial x_j$ of Equation~(\ref{eq:NFWP_lens_equation}) gives us the Jacobian
\begin{multline}
\mathrm{det}\,J(\boldsymbol x)= \left\{ 1-\frac{4\,\kappa_{\text{s}}}{x^2}\, \left[\ln{\frac{x}{2}}+\mathcal{F}(x)\right] -\frac{\kappa_{\text{P}}}{|\boldsymbol x - \bm{x_{\text{P}}}|^2} \right\}\\ \times \left\{ 1+\frac{4\,\kappa_{\text{s}}}{x^2}\, \left[\ln{\frac{x}{2}}+\mathcal{F}(x)\right] +\frac{\kappa_{\text{P}}}{|\boldsymbol x - \bm{x_{\text{P}}}|^2} -4\,\kappa_{\text{s}}\,\frac{\mathcal{F}(x)-1}{1-x^2} \right\}\\ +16\,\kappa_{\text{s}}\,\kappa_{\text{P}}\,\frac{|\,\bm{x}\times \bm{x_{\text{P}}}\,|^2}{x^4|\boldsymbol x - \bm{x_{\text{P}}}|^4}\, \left\{ \ln{\frac{x}{2}}+\mathcal{F}(x)-\frac{x^2}{2}\,\frac{\mathcal{F}(x)-1}{1-x^2} \right\}\,,
\label{eq:NFWP_Jacobian}
\end{multline}
written here in a form independent of coordinate-frame orientation. In the orientation with the point mass on the positive horizontal axis, the norm of the cross product $|\,\bm{x}\times \bm{x_{\text{P}}}\,|=x_{\text{P}}\,|x_2|$.

Far from the center of the halo ($x=0$) and far from the point mass ($\bm{x}=\bm{x_{\text{P}}}$) the Jacobian $\mathrm{det}\,J\to 1$. In the general case, with $x_{\text{P}}\neq 0$, the Jacobian has two divergences. As shown in Section~\ref{sec:NFW-Jacobian}, at the center of the halo $\mathrm{det}\,J(\boldsymbol x)\approx4\,\kappa_{\text{s}}^2\,\ln^2{x}\to\infty$, while at the position of the point mass $\mathrm{det}\,J(\boldsymbol x)\approx-\kappa_{\text{P}}^2/|\boldsymbol x - \bm{x_{\text{P}}}|^4\to-\infty$.

\subsection{Point mass at the halo center}
\label{sec:central_position}

In the special case when the point mass is positioned at the center of the halo ($x_{\text{P}}=0$) the Jacobian loses the entire final term in Equation~(\ref{eq:NFWP_Jacobian}), leaving the factorized part
\begin{multline}
\mathrm{det}\,J(\boldsymbol x)= \left\{ 1-\frac{4\,\kappa_{\text{s}}}{x^2}\, \left[\ln{\frac{x}{2}}+\mathcal{F}(x)\right] -\frac{\kappa_{\text{P}}}{x^2} \right\}\\ \times \left\{ 1+\frac{4\,\kappa_{\text{s}}}{x^2}\, \left[\ln{\frac{x}{2}}+\mathcal{F}(x)\right] +\frac{\kappa_{\text{P}}}{x^2} -4\,\kappa_{\text{s}}\,\frac{\mathcal{F}(x)-1}{1-x^2} \right\}\,.
\label{eq:NFWP_Jacobian_central}
\end{multline}
Here the situation is peculiar, since the two opposite divergences coincide at the origin. The stronger one due to the point mass prevails, so that $\mathrm{det}\,J(\boldsymbol x)\to-\infty$ as $x\to  0$.

The factor in the first braces in Equation~(\ref{eq:NFWP_Jacobian_central}) yields the tangential critical curve, which is a circle $|\boldsymbol x|=x_{\text{PT}}$ with radius obtained by numerically solving
\beq
1-\frac{4\,\kappa_{\text{s}}}{x_{\text{PT}}^2}\, \left[\,\ln{\frac{x_{\text{PT}}}{2}}+\mathcal{F}(x_{\text{PT}})\,\right] -\frac{\kappa_{\text{P}}}{x_{\text{PT}}^2}=0\,.
\label{eq:tangential_central}
\eeq
The equation has a single solution for any combination of $\kappa_{\text{s}}$ and $\kappa_{\text{P}}$. Substituting the critical curve $\boldsymbol x=x_{\text{PT}}\,(\cos{\varphi},\sin{\varphi})$ with $\varphi\in [\,0, 2\pi\,]$ in Equation~(\ref{eq:NFWP_lens_equation}) and using Equation~(\ref{eq:tangential_central}), we obtain the corresponding part of the caustic:
\beq
\boldsymbol y = \boldsymbol x\,\left\{1-\frac{4\,\kappa_{\text{s}}}{x_{\text{PT}}^2}\, \left[\,\ln{\frac{x_{\text{PT}}}{2}}+\mathcal{F}(x_{\text{PT}})\,\right]-\frac{\kappa_{\text{P}}}{x_{\text{PT}}^2}\right\}=(\,0,0\,)\,.
\label{eq:tangential_caustic_central}
\eeq
The tangential part of the caustic remains unchanged, consisting of the single point at the origin.

Setting the second braces in Equation~(\ref{eq:NFWP_Jacobian_central}) equal to zero yields the radial critical-curve equation. The solutions are circles with radius $x_{\text{PR}}$ obtained by numerically solving
\beq
1+\frac{4\,\kappa_{\text{s}}}{x_{\text{PR}}^2}\, \left[\,\ln{\frac{x_{\text{PR}}}{2}}+\mathcal{F}(x_{\text{PR}})\,\right] +\frac{\kappa_{\text{P}}}{x_{\text{PR}}^2} -4\,\kappa_{\text{s}}\,\frac{\mathcal{F}(x_{\text{PR}})-1}{1-x_{\text{PR}}^2} =0\,.
\label{eq:radial_central}
\eeq
However, here the number of solutions for a given value of $\kappa_{\text{s}}$ depends on the value of $\kappa_{\text{P}}$. Substituting the radial critical curve $\boldsymbol x=x_{\text{PR}}\,(\cos{\varphi},\sin{\varphi})$ with $\varphi\in [\,0, 2\pi\,]$ in Equation~(\ref{eq:NFWP_lens_equation}) and using Equation~(\ref{eq:radial_central}), we obtain the corresponding part of the caustic:
\begin{multline}
\boldsymbol y = \boldsymbol x\,\left\{1-\frac{4\,\kappa_{\text{s}}}{x_{\text{PR}}^2}\, \left[\,\ln{\frac{x_{\text{PR}}}{2}}+\mathcal{F}(x_{\text{PR}})\,\right]-\frac{\kappa_{\text{P}}}{x_{\text{PR}}^2}\right\} \\ =-2\,\left[\,2\,\kappa_{\text{s}}\,\frac{\mathcal{F}(x_{\text{PR}})-1}{1-x_{\text{PR}}^2}-1\,\right]\, x_{\text{PR}}\,(\cos{\varphi},\sin{\varphi})\,,
\label{eq:radial_caustic_central}
\end{multline}
where the expression in the square brackets is equal to $\kappa(x_{\text{PR}})-1$, which is positive. For each radial critical curve, the corresponding part of the caustic is thus a circle with radius
\beq
y_{\text{PR}}=2\,x_{\text{PR}}\,[\,\kappa(x_{\text{PR}})-1\,]\,.
\label{eq:yR_central}
\eeq

\bfi
{\centering
\vspace{0cm}
\hspace{-0.5cm}
\includegraphics[width=6 cm]{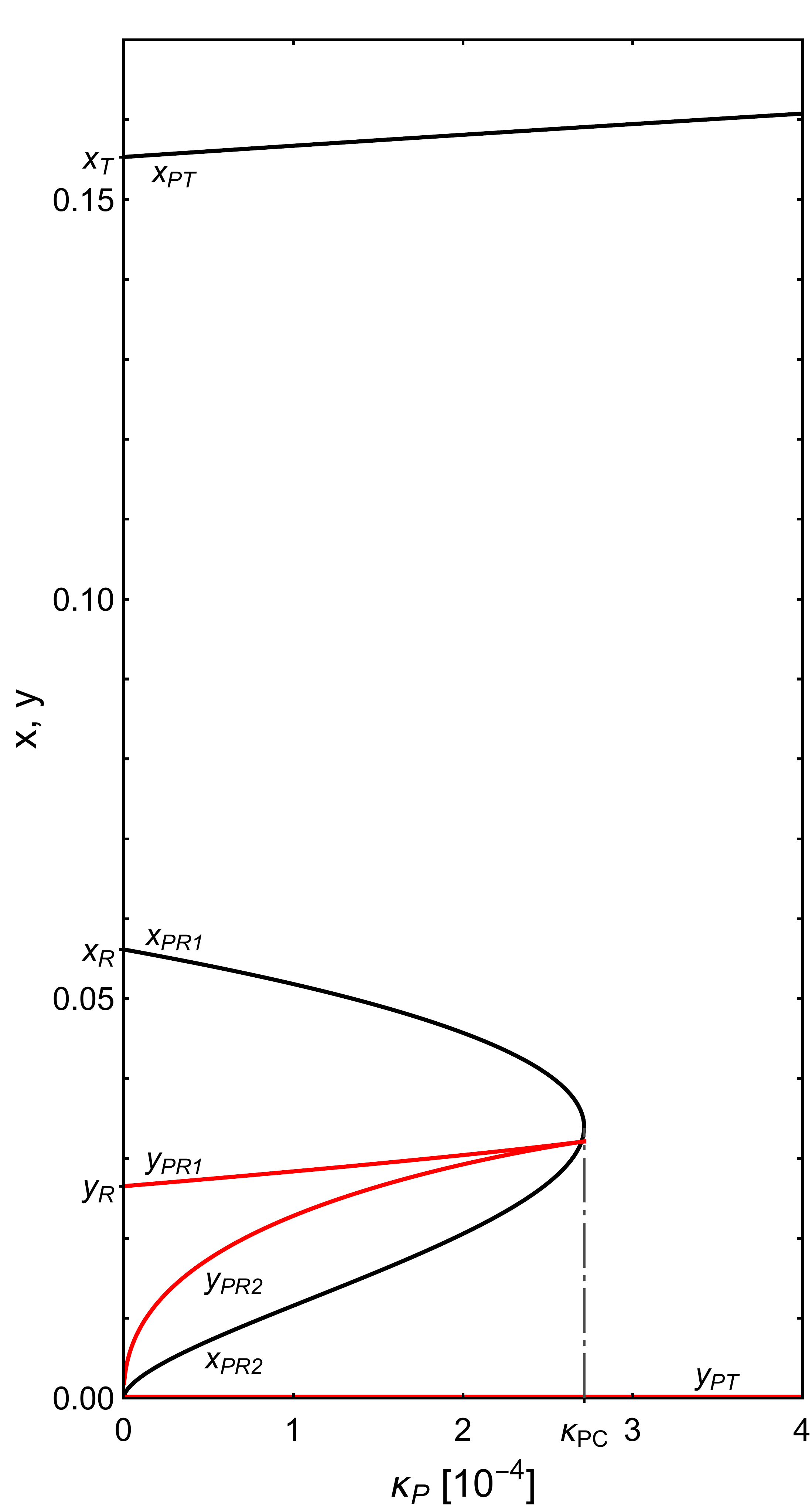}
\caption{Radii of critical curves (black: tangential $x_{\text{PT}}$, radial $x_{\text{PR1}}$ and $x_{\text{PR2}}$) and caustics (red: tangential $y_{\text{PT}}=0$, radial $y_{\text{PR1}}$ and $y_{\text{PR2}}$) of a $\kappa_{\text{s}}\approx 0.239035$ NFW halo density profile with a centrally positioned point mass, plotted as a function of its mass parameter $\kappa_{\text{P}}$. The corresponding radii $x_{\text{T}},\, x_{\text{R}},\, y_{\text{R}}$ of an unperturbed NFW halo (intersections with the dot-dashed line in Figure~\ref{fig:NFW_radii}) are marked along the vertical axis. Note the vanishing of the radial critical curves and caustics at $\kappa_{\text{P}}=\kappa_{\text{PC}}\approx 2.714 \cdot 10^{-4}$ (marked by the dot-dashed line).}
\label{fig:NFWP_radii}}
\efi

We present the structure of the critical curves and caustics in Figure~\ref{fig:NFWP_radii} as a function of $\kappa_{\text{P}}$ for the cluster with $\kappa_{\text{s}}\approx 0.239035$ chosen for illustration in Figure~\ref{fig:NFW}. The radius of the tangential critical curve $x_{\text{PT}}$ is a simple monotonically increasing function of $\kappa_{\text{P}}$, which starts on the vertical axis at $x_{\text{T}}$, the radius of the NFW tangential critical curve. This growth is just a simple consequence of the increasing mass at the origin.

More interesting are the radial critical curves. For low values of $\kappa_{\text{P}}$ Equation~(\ref{eq:radial_central}) has two solutions. The larger one ($x_{\text{PR1}}$) starts at $x_{\text{R}}$, the radius of the NFW radial critical curve, and decreases with $\kappa_{\text{P}}$. The smaller one ($x_{\text{PR2}}$) starts at 0 and increases with $\kappa_{\text{P}}$. As $\kappa_{\text{P}}$ grows, the two curves approach each other until they merge at a critical value $\kappa_{\text{P}}=\kappa_{\text{PC}}\approx 2.714 \cdot 10^{-4}$ (for our choice of $\kappa_{\text{s}}$). For super-critical $\kappa_{\text{P}}>\kappa_{\text{PC}}$ Equation~(\ref{eq:radial_central}) has no solution, i.e., there are no radial critical curves.

The two corresponding radial components of the caustic reflect the behavior of the critical curves, with $y_{\text{PR1}}$ starting at $y_{\text{R}}$, the radius of the NFW radial caustic, and $y_{\text{PR2}}$ starting at 0. Even though the larger radial caustic grows with $\kappa_{\text{P}}$, the two caustic circles approach each other faster than the corresponding critical-curve circles and vanish beyond $\kappa_{\text{PC}}$.

The critical curves and caustics in the two regimes are illustrated by the panels in the bottom row of Figure~\ref{fig:gallery-print}. The two left columns show the Jacobian plot with critical curves and the total magnification map with caustics in the sub-critical case (for $\kappa_{\text{P}}=10^{-4}$), the two right columns show the corresponding plots in the super-critical case (for $\kappa_{\text{P}}=10^{-3}$). The notation is the same as in Figure~\ref{fig:NFW}, with the additional cyan circle indicating the position and Einstein ring of the point mass in units of the angular scale radius, $\theta_{\text{EP}}\,D_{\text{l}}/r_{\text{s}}\,$.

For the lower-mass case with $\kappa_{\text{P}}=10^{-4}$, the Jacobian plot resembles the plot from Figure~\ref{fig:NFW}, except the region near the origin. In this example the additional radial critical curve is marginally larger than the point-mass Einstein radius. Within it the point mass dominates and $\mathrm{det}\,J<0$. Between the two radial critical curves $\mathrm{det}\,J>0$, between the outer radial and tangential critical curves $\mathrm{det}\,J<0$, followed by $\mathrm{det}\,J>0$ outside the tangential critical curve. Instead of the single radial caustic of the NFW halo seen in Figure~\ref{fig:NFW}, there are two circular radial caustics, with the outer and inner one corresponding to the outer and inner radial critical curve, respectively. The total magnification diverges at both caustics from the side of the annular region between them.

As $\kappa_{\text{P}}$ increases, so does the negative-Jacobian region around the origin dominated by the point mass, while the outer radial critical curve and the enclosed positive-Jacobian annulus shrink. The outer radial caustic grows, but the inner one grows faster. At $\kappa_{\text{P}}=\kappa_{\text{PC}}$ the two radial critical curves (as well as the two radial caustics) merge, causing the positive-Jacobian annulus to vanish. This situation is illustrated in the bottom panels of the two central columns of Figure~\ref{fig:gallery-online}. The peculiar properties of the single merged radial critical curve and caustic are described in Appendix~\ref{sec:Appendix-vanishing_curves}. For larger $\kappa_{\text{P}}$ the radial critical curves and caustics disappear.

For the higher-mass case with $\kappa_{\text{P}}=10^{-3}$, only the tangential critical curve remains and the negative-Jacobian region around the origin extends all the way to it, as shown in Figure~\ref{fig:gallery-print}. The caustic is reduced to the single point at the origin, while the magnification shows a brighter central region roughly the size of the vanished radial caustic.

Comparison with Figure~\ref{fig:NFW} illustrates how dramatically a single added object may change the lensing properties of the NFW halo. The critical value $\kappa_{\text{PC}}\approx 2.714 \cdot 10^{-4}$ for our halo parameter choice given by Equation~(\ref{eq:parameter_choice}) can be used in Equation~(\ref{eq:kappa_P}) to compute the corresponding critical mass ratio $M_{\text{P}}/M_{\text{NFW}}\approx 3.91 \cdot 10^{-4}$. Even though such a relative mass may seem low, its gravitational field is strong enough to destroy the radial critical curve and caustic of the NFW halo in which it is embedded.

Similar behavior has been found previously for a central point mass embedded in different axially symmetric mass distributions, such as in a cored isothermal sphere \citep{mao_etal01} or in a Plummer model \citep{werner_evans06}. In either of these models there is a critical mass of the central point, above which the lens has no radial critical curves. Nevertheless, this behaviour is not ubiquitous: a central point mass embedded in a singular isothermal sphere has no radial critical curves, irrespective of its mass \citep{mao_witt12}.

\subsection{Point mass at a general position}
\label{sec:general_position}

When positioning the point mass at increasing distances $x_{\text{P}}$ from the halo center, the critical curve undergoes a sequence of transitions in which its loops connect, disconnect, appear, or vanish. These transitions are accompanied by underlying metamorphoses of the caustic \citep{schneider_etal92}. The curves start from the central configurations described in Section~\ref{sec:central_position} and end with separate critical curves and caustics of a NFW halo and a distant point-mass lens. In Section~\ref{sec:galleries} we describe the overall changes of the critical curves and caustics. In Section~\ref{sec:boundaries} we explore the parameter space of the point mass and identify boundaries at which the critical-curve transitions occur. We illustrate the details of several transitions and transition sequences in Section~\ref{sec:transition_details}.

\subsubsection{Critical-curve and caustic galleries}
\label{sec:galleries}

\renewcommand{\thefigure}{5.A}
\begin{figure*}
{\centering
\vspace{0cm}
\hspace{0cm}
\includegraphics[width=16 cm]{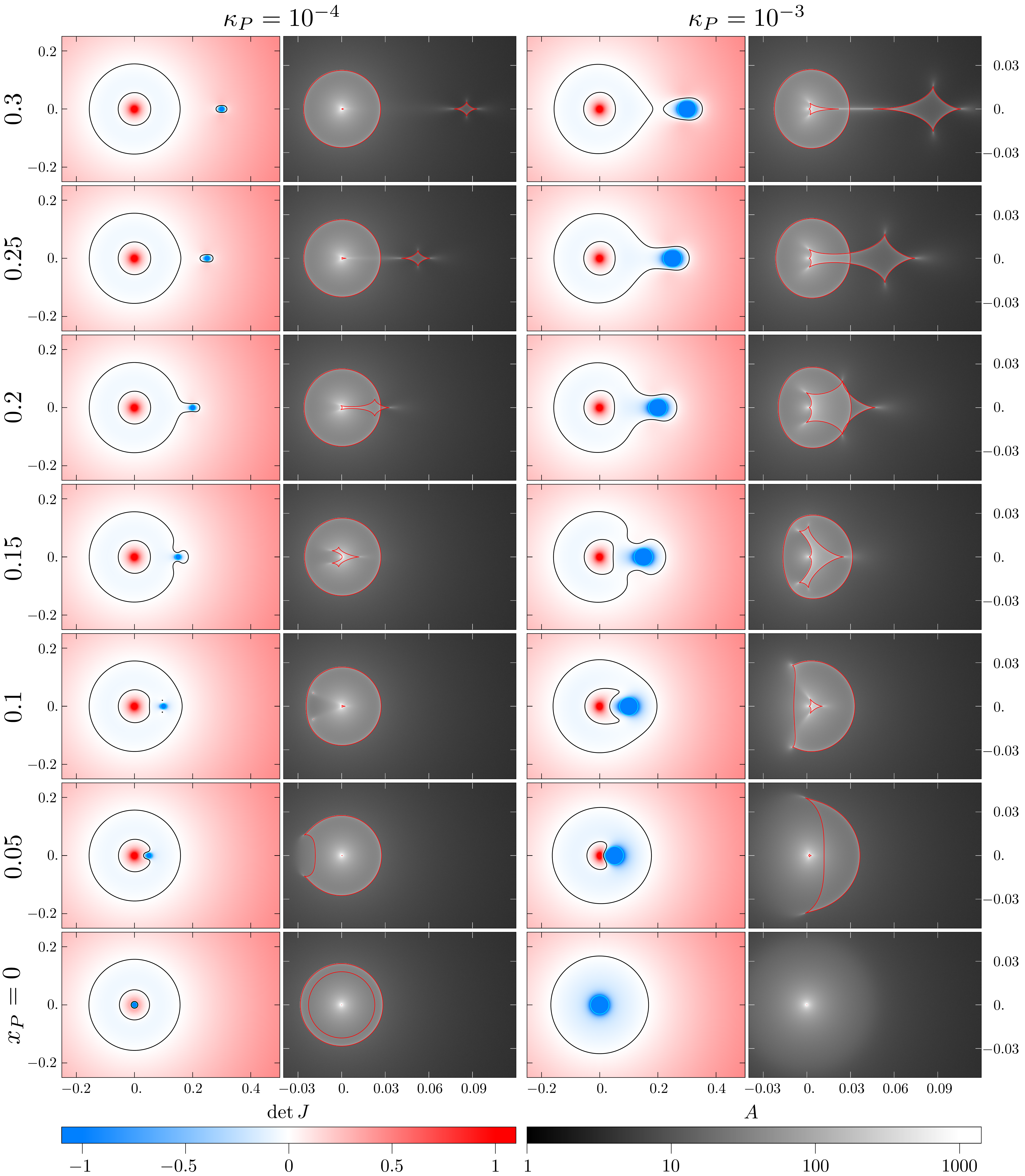}
\caption{Critical curves and caustics of a NFW halo + point-mass lens. The left pair of columns corresponds to a sub-critical mass parameter $\kappa_{\text{P}}=10^{-4}$ and the right pair to a super-critical $\kappa_{\text{P}}=10^{-3}$, as marked at the top. The rows correspond to positions of the point mass ranging from $x_{\text{P}}=0$ to $x_{\text{P}}=0.3$, as marked along the left side. The point-mass parameter-combination grid is marked by red crosses in Figure~\ref{fig:transitions}. The cyan circles indicate the point-mass position and its Einstein ring. Here $\mathrm{det}\,J\to-\infty$ at the position of the point mass, and $\mathrm{det}\,J\to\infty$ at the origin (except when the point mass lies there). Remaining notation and color bars same as in Figure~\ref{fig:NFW}.}
\label{fig:gallery-print}}
\end{figure*}

\renewcommand{\thefigure}{5.B}
\begin{figure*}
{\centering
\vspace{0cm}
\hspace{0cm}
\includegraphics[height=20 cm]{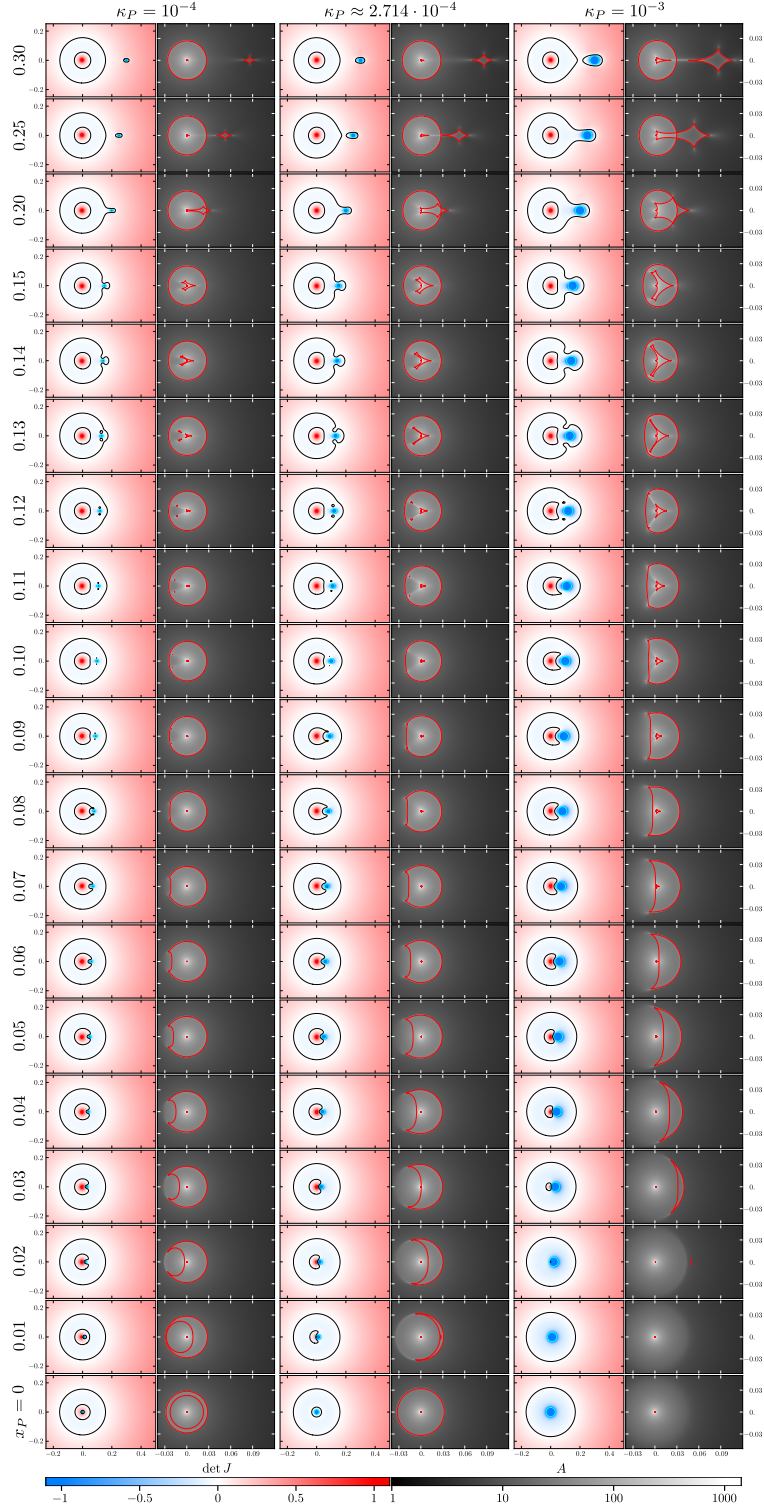}
\caption{Critical curves and caustics of a NFW halo + point-mass lens. The left pair of columns corresponds to a sub-critical mass parameter $\kappa_{\text{P}}=10^{-4}$, the central pair to critical $\kappa_{\text{P}}=\kappa_{\text{PC}}\approx 2.714 \cdot 10^{-4}$, and the right pair to super-critical $\kappa_{\text{P}}=10^{-3}$, as marked at the top. The rows correspond to positions of the point mass ranging from $x_{\text{P}}=0$ to $x_{\text{P}}=0.3$, as marked along the left side. The point-mass parameter-combination grid is marked by red and black crosses in Figure~\ref{fig:transitions}. Notation same as in Figure~\ref{fig:gallery-print}.}
\label{fig:gallery-online}}
\end{figure*}

In Figure~\ref{fig:gallery-print} we present a sample critical-curve and caustic gallery for a sub-critical ($\kappa_{\text{P}}=10^{-4}$, left columns) and a super-critical ($\kappa_{\text{P}}=10^{-3}$, right columns) point mass. Its positions are marked in the critical-curve plots by the cyan Einstein-radius circles on the horizontal axis. The radial distances from the origin are indicated along the left side of the figure, increasing from $x_{\text{P}}=0$ in the bottom row to $x_{\text{P}}=0.3$ in the top row, in regular steps of $0.05$.

Going up from $x_{\text{P}}=0$ to $x_{\text{P}}=0.05$ in the sub-critical case, the small critical-curve loop around the point mass connects and merges with the perturbed radial critical curve of the NFW halo. The inner circular caustic connects and merges at its left side with the perturbed radial NFW caustic in a beak-to-beak metamorphosis, while its right side progresses across the origin to the left. The four-cusped perturbed tangential NFW caustic is too tiny to be distinguished from a point at the origin.

By $x_{\text{P}}=0.1$ two small critical-curve loops have detached from the perturbed radial NFW critical curve. On the caustic, two small three-cusped loops have detached in simultaneous beak-to-beak metamorphoses from the perturbed radial NFW caustic. Details of this transition are described in Section~\ref{sec:transition_details} further below. Between $x_{\text{P}}=0.1$ and $x_{\text{P}}=0.15$, the two small critical-curve loops connect and merge with the perturbed tangential NFW critical curve. The two small three-cusped caustic loops connect and merge with the perturbed tangential NFW caustic in simultaneous beak-to-beak metamorphoses, forming a central six-cusped caustic loop.

For a more distant position of the point mass, a small critical-curve loop detaches from the perturbed tangential NFW critical curve, as seen for $x_{\text{P}}=0.25$. The small loop eventually converges to the point-mass Einstein ring. The horizontally stretched six-cusped caustic loop seen at $x_{\text{P}}=0.2$ disconnects in a beak-to-beak metamorphosis, forming two four-cusped loops. The smaller central one eventually shrinks to the point-like tangential NFW caustic. With increasing $x_{\text{P}}$, the larger one comes to resemble the four-cusped Chang--Refsdal caustic of a point mass with a constant low external shear \citep{chang_refsdal84}. Eventually, it shrinks to the point-like caustic of an isolated point-mass lens.

The influence of the point mass on the critical curve and caustic in the super-critical case is more pronounced and better visible than in the sub-critical case. Nevertheless, for both mass parameters we see that the critical curve is affected in a region reaching multiple Einstein radii from the point mass.

The initial transitions are markedly different in the super-critical case. As described in Section~\ref{sec:central_position}, the critical curve for $x_{\text{P}}=0$ consists only of the perturbed tangential NFW critical curve. Nevertheless, a perturbed radial NFW critical curve appears already by $x_{\text{P}}=0.05$. The perturbed tangential NFW caustic evolves from a single point at the origin for $x_{\text{P}}=0$ to a tiny four-cusped loop for $x_{\text{P}}=0.05$. In addition, another caustic loop with two cusps has appeared by $x_{\text{P}}=0.05$ in a lips metamorphosis, forming a crescent-shaped rather strongly perturbed radial NFW caustic. A detailed description of the super- and sub-critical transition sequences for point-mass positions close to the halo center is presented below in Section~\ref{sec:transition_details}.

The subsequent transitions are similar to the sub-critical case, though some of them are not seen in Figure~\ref{fig:gallery-print}. This is the case for the two small loops detaching from the perturbed radial NFW critical curve \& caustic and connecting to the perturbed tangential NFW critical curve \& caustic, which occur here between the $x_{\text{P}}=0.1$ and $x_{\text{P}}=0.15$ rows. These can be seen in Figure~\ref{fig:gallery-online}, which shows a more detailed gallery of critical curves and caustics. In addition to the positions and masses from Figure~\ref{fig:gallery-print} it includes positions from $x_{\text{P}}=0$ to $x_{\text{P}}=0.15$ in a finer step of $0.01$, as well as panels for the critical mass parameter $\kappa_{\text{PC}}\approx 2.714 \cdot 10^{-4}$ in the central two columns. For this intermediate mass, note in the $x_{\text{P}}=0.01$ panels the additional tiny critical-curve loop close to the origin and the corresponding two-cusped caustic loop inside the larger caustic crescent.

More generally, all individual transitions occur at different radial positions for the different mass parameters. For example, the detachment of the two small critical-curve loops from the perturbed radial NFW critical curve occurs between $x_{\text{P}}=0.08$ and $x_{\text{P}}=0.09$ for $\kappa_{\text{P}}=10^{-4}$, between $x_{\text{P}}=0.09$ and $x_{\text{P}}=0.1$ for $\kappa_{\text{P}}=\kappa_{\text{PC}}$, and between $x_{\text{P}}=0.11$ and $x_{\text{P}}=0.12$ for $\kappa_{\text{P}}=10^{-3}$. We illustrate this in detail in the following Section~\ref{sec:boundaries}, where we map the occurrence of all critical-curve transitions in the parameter space of the point mass.

\subsubsection{Boundaries in point-mass parameter space}
\label{sec:boundaries}

We track the changing structure of critical curves and caustics in the point-mass parameter-space region defined by the intervals $\kappa_{\text{P}}\in[0,\,0.0035]$ and $x_{\text{P}}\in[0,\,0.4]$. Within this space we identify boundaries at which the overall topology of the critical curve changes in transitions such as those seen in Section~\ref{sec:galleries}. These involve situations in which critical-curve loops connect/disconnect, or appear/vanish. In addition, we also identify situations in which loops shrink to a point, or touch without connecting. We chose the upper limits on the parameter intervals empirically: the $x_{\text{P}}$ limit lies above the final transition in the studied $\kappa_{\text{P}}$ interval; for values above the $\kappa_{\text{P}}$ limit there are no further changes to the structure of the boundaries.

We note that straightforward use of the first expression for $\mathcal{F}(x)$ from Equation~(\ref{eq:f(x)}) leads to numerical instabilities for $x\ll 1$, close to the origin of the image plane. For tracking the changing structure of critical curves in this region it was necessary to use the exact and stable expressions given by Equation~(\ref{eq:f(x)_equivalent}) and Equation~(\ref{eq:ln_f(x)_equivalent}), as shown in Appendix~\ref{sec:Appendix-origin}.

\renewcommand{\thefigure}{6}
\begin{figure*}
{\centering
\begin{tabular}{cc}
  \begin{tabular}{c}
    \\
    \vspace{-0.1cm}
    \hspace{-1cm}\includegraphics[scale=0.552]{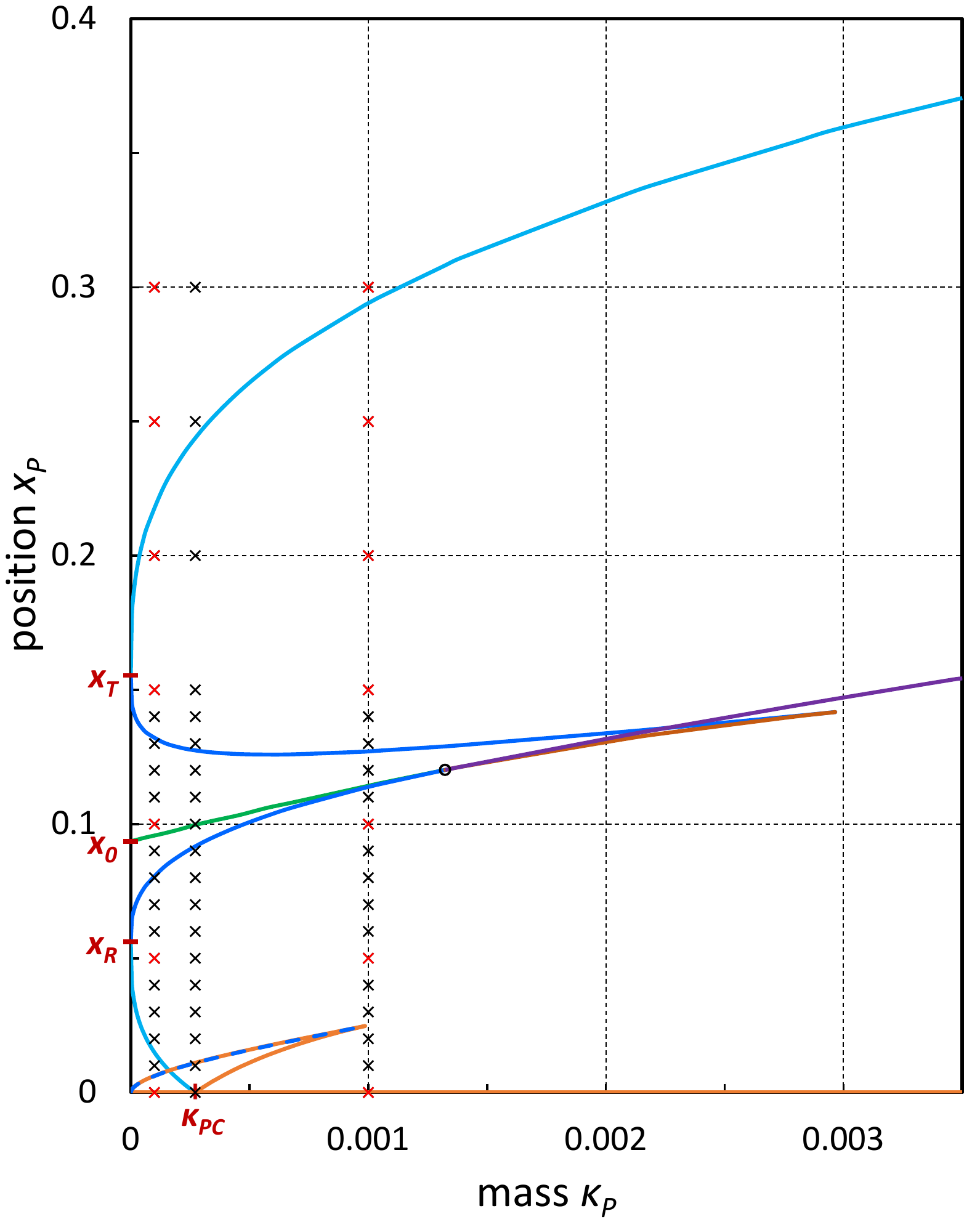} \\
  \end{tabular}
  &
  \begin{tabular}{l}
    \\
    \vspace{0.25cm}
    \hspace{-1.0cm}\includegraphics[scale=0.5016]{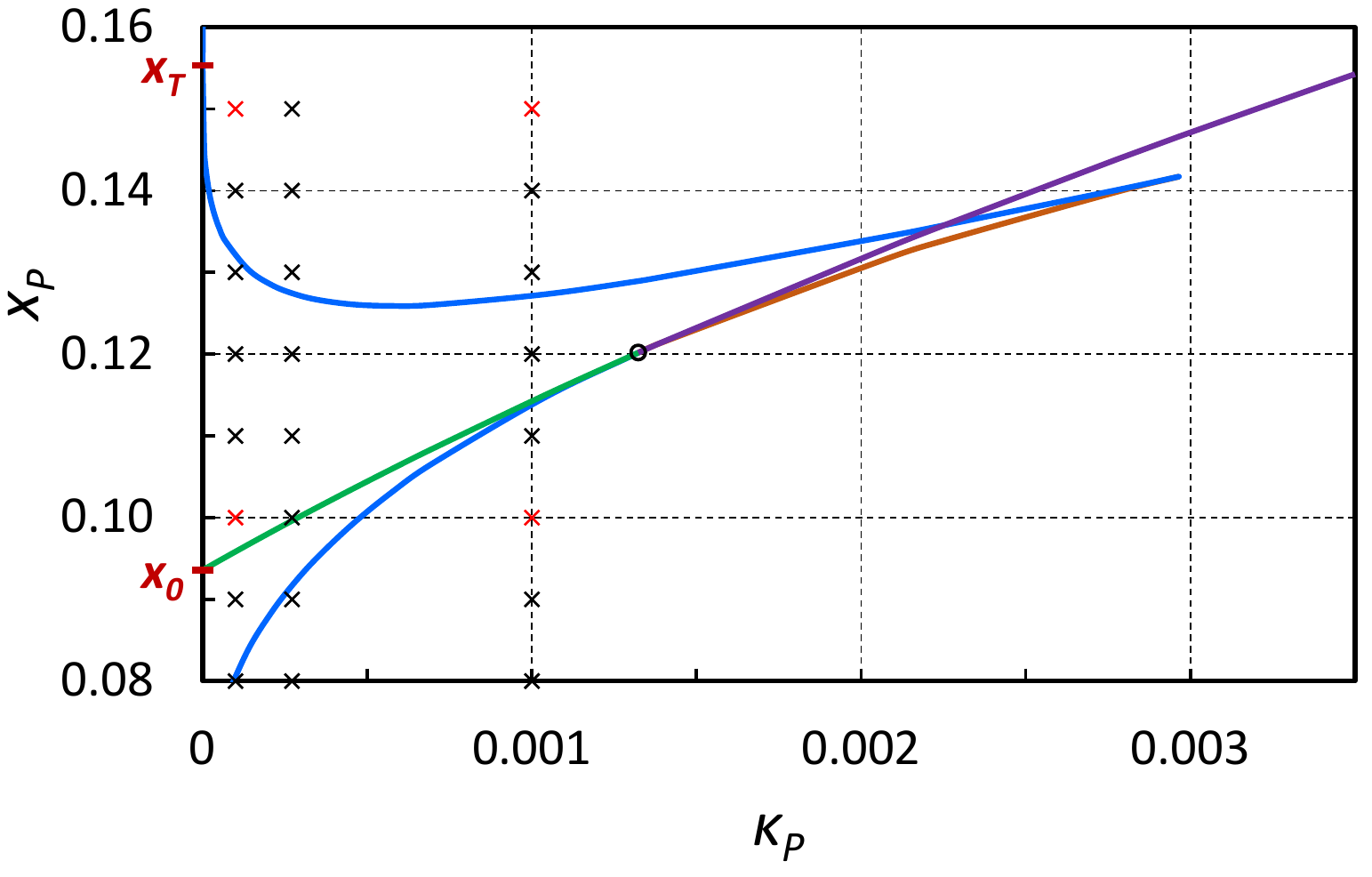} \\
    \vspace{0.2cm}
    \\
    \hspace{-1.0cm}\includegraphics[scale=0.495]{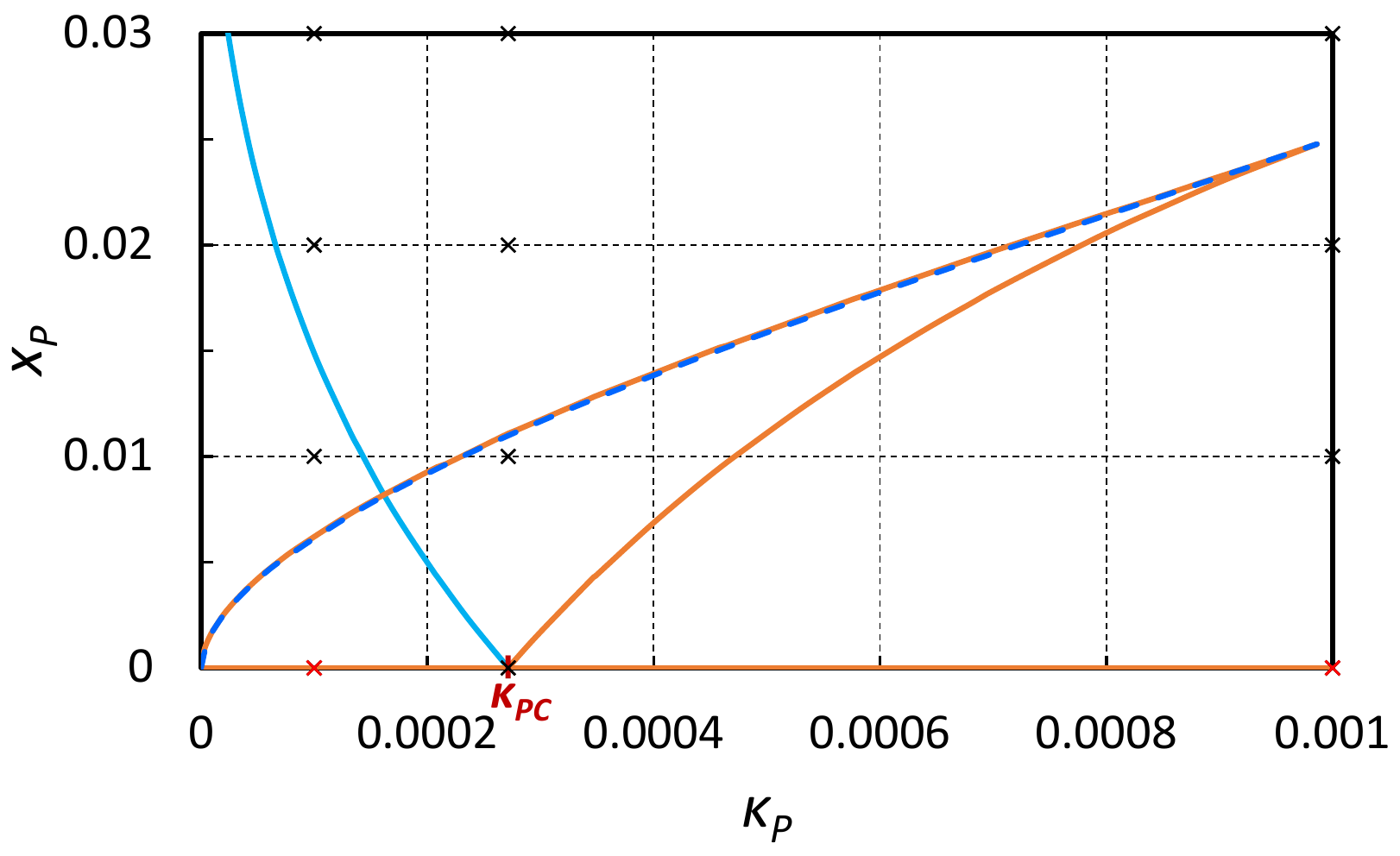}
  \end{tabular}
\end{tabular}
\caption{Critical-curve transitions in the position $x_{\text{P}}$ vs. mass parameter $\kappa_{\text{P}}$ parameter space of a point mass in a NFW halo. The transitions occur at boundaries colored according to the corresponding caustic metamorphoses: single beak-to-beak (cyan), two beak-to-beaks (blue), single lips (orange, including the horizontal axis), two lips (dark orange), two elliptic umbilics (green), two hyperbolic umbilics (violet), two parabolic umbilics (point marked by black circle). The blue/orange dashed line at bottom left indicates two closely adjacent transitions: the orange lips occur at larger $x_{\text{P}}$ than the dashed blue beak-to-beaks. Crosses indicate the parameter combinations of the examples illustrated in Figure~\ref{fig:gallery-online}; red crosses mark the examples illustrated in Figure~\ref{fig:gallery-print}. Top right panel: detail of the intermediate-position transitions near the umbilics; bottom right panel: detail of the transitions close to the origin. Additional ticks on the axes mark the critical mass parameter $\kappa_{\text{PC}}\approx 2.714 \cdot 10^{-4}$, the NFW halo tangential ($x_{\text{T}}$) and radial ($x_{\text{R}}$) critical-curve radii, and the unit-convergence radius $x_0$.}
\label{fig:transitions}}
\end{figure*}

Finding the boundaries involved several steps. We started by computing the critical curves and caustics on a rough parameter grid, inspecting the results to identify pairs of neighboring grid points with different topologies of the critical curve and different caustic structures. For each such pair we proceeded by interval-halving to pinpoint the intersection of the boundary between the points with the respective grid-line. Where necessary, we added more points in between these grid-line intersections to obtain smoother boundaries. In the emerging boundary plot we checked all mass-parameter intervals with different vertical sequences of boundaries to make sure the changes across the boundaries agreed with the characteristics of the corresponding metamorphoses (such as characteristic changes in numbers of loops, or changes in numbers of cusps on caustic loops). Finally, we checked the continuity of the critical curves and caustics close to the axes, which correspond to analytically studied axisymmetric lenses: in the zero-mass limit close to the vertical axis (comparison with Section~\ref{sec:NFW-Jacobian} and Figure~\ref{fig:NFW}), and in the zero-displacement limit close to the horizonal axis (comparison with Section~\ref{sec:central_position} and Figure~\ref{fig:gallery-online}).

The mapped boundaries are presented in a $x_{\text{P}}$ vs. $\kappa_{\text{P}}$ plot in Figure~\ref{fig:transitions}. The left panel shows the full explored parameter space; the right panels show two expanded regions in more detail: bottom right for low masses close to the origin; top right for intermediate positions. For better interpretation of the results, the red crosses indicate the parameter combinations for which critical curves and caustics are shown in Figure~\ref{fig:gallery-print}; the critical curves and caustics for the full set of red and black crosses are shown in Figure~\ref{fig:gallery-online}. For additional orientation, we mark the NFW halo radial and tangential critical-curve radii ($x_{\text{R}}$ and $x_{\text{T}}$, respectively) on the vertical axis, and the critical mass parameter $\kappa_{\text{PC}}$ on the horizontal axis.

The colors of the boundaries indicate the type and multiplicity of the associated caustic metamorphosis. Single metamorphoses occur in the source plane at a point on the symmetry axis passing through the point mass and the halo center, while two same simultaneous metamorphoses occur at points symmetrically offset from the axis. Cyan color indicates a single beak-to-beak metamorphosis, occurring here at two boundaries: from $x_{\text{R}}$ on the vertical axis downward to $\kappa_{\text{PC}}$ on the horizontal axis; from $x_{\text{T}}$ on the vertical axis upward. Blue indicates two simultaneous beak-to-beak metamorphoses, occurring here at three boundaries: from the origin up to the boundary spike just under $\kappa_{\text{P}}=0.001$ (plotted by a dashed line to reveal the closely adjacent orange lips boundary); from $x_{\text{R}}$ on the vertical axis upward to the point marked by the circle; from $x_{\text{T}}$ on the vertical axis first downward, then rising to the boundary spike just under $\kappa_{\text{P}}=0.003$.

Orange indicates a single lips metamorphosis, occurring here at three boundaries: from the origin along the entire horizontal axis (corresponding to $x_{\text{P}}=0$); from the origin upward to the boundary spike just under $\kappa_{\text{P}}=0.001$ (just above the closely adjacent dashed blue boundary); from $\kappa_{\text{PC}}$ on the horizontal axis upward to the same boundary spike. Dark orange indicates two simultaneous lips metamorphoses, occurring here at one boundary: from the point marked by the circle upward to the boundary spike just under $\kappa_{\text{P}}=0.003$. Green indicates two simultaneous elliptic umbilic metamorphoses, occurring here at one boundary: from the unit-convergence radius $x_0\approx 0.0936$ on the vertical axis upward to the point marked by the circle. Violet indicates two simultaneous hyperbolic umbilic metamorphoses, occurring here at one boundary: from the point marked by the circle upward. Finally, the point marked by the black circle indicates two simultaneous parabolic umbilics. Four boundaries meet at this point, all of them corresponding to two simultaneous metamorphoses. In clockwise order from lower left these are beak-to-beak, elliptic umbilic, hyperbolic umbilic, and lips boundaries.

In addition to the metamorphoses for which we plotted the parameter-space boundaries in Figure~\ref{fig:transitions}, the caustic undergoes also the swallow-tail metamorphosis. In it the caustic develops a local ``twist'' with a self-crossing point and two additional cusps. However, unlike the plotted metamorphoses, the swallow tail has no effect on the critical-curve topology. It occurs here always simultaneously in twos, always closely adjacent to simultaneous beak-to-beak boundaries. One such simultaneous swallow-tail pair occurs between the beak-to-beak and lips boundaries extending in Figure~\ref{fig:transitions} from the origin upward to the boundary spike just under $\kappa_{\text{P}}=0.001$; two such simultaneous pairs occur in close succession just beneath the beak-to-beak boundary extending from $x_{\text{R}}$ on the vertical axis upward to the point marked by the circle.

Before exploring individual transition sequences, metamorphoses, and boundaries in more detail in Section~\ref{sec:transition_details}, it is worth pointing out the astounding richness of structure, transitions, and lensing regimes between them, which was quite unexpected for us in such a simple lens model.

\subsubsection{Details of specific transitions and transition sequences}
\label{sec:transition_details}

The first interesting transition occurs along the horizontal axis of the parameter-space plot in Figure~\ref{fig:transitions}, i.e., for an arbitrarily small displacement of the point mass from the halo center. In such a situation the two opposite Jacobian divergences described in Section~\ref{sec:NFWP-Jacobian} no longer coincide, and the Jacobian must cross zero at some point between them. This results in a small new critical-curve loop encircling the weaker $\mathrm{det}\,J(\boldsymbol x)\to\infty$ divergence at the origin of the image plane. The caustic undergoes a lips metamorphosis, in which a new two-cusped loop forms beyond the point-mass position. This particular lips metamorphosis is unusual in its appearance at infinity rather than at a finite distance from the origin of the source plane. A further increase in the point-mass displacement brings the new caustic loop rapidly toward the origin.

Inspecting the changes along the left column of crosses in Figure~\ref{fig:transitions} corresponding to the sub-critical $\kappa_{\text{P}}=10^{-4}$ example, we see that the critical curve undergoes a total of eight transitions as we vary the point-mass position from the halo center to an asymptotic distance. Only a few of these can be identified in the galleries in Figure~\ref{fig:gallery-print} and Figure~\ref{fig:gallery-online}, as described in Section~\ref{sec:galleries}. In order to provide a more comprehensive overview we present the details of three transition sequences below.

\renewcommand{\thefigure}{7}
\bfi
{\centering
\vspace{0cm}
\hspace{0cm}
\includegraphics[width=8.5 cm]{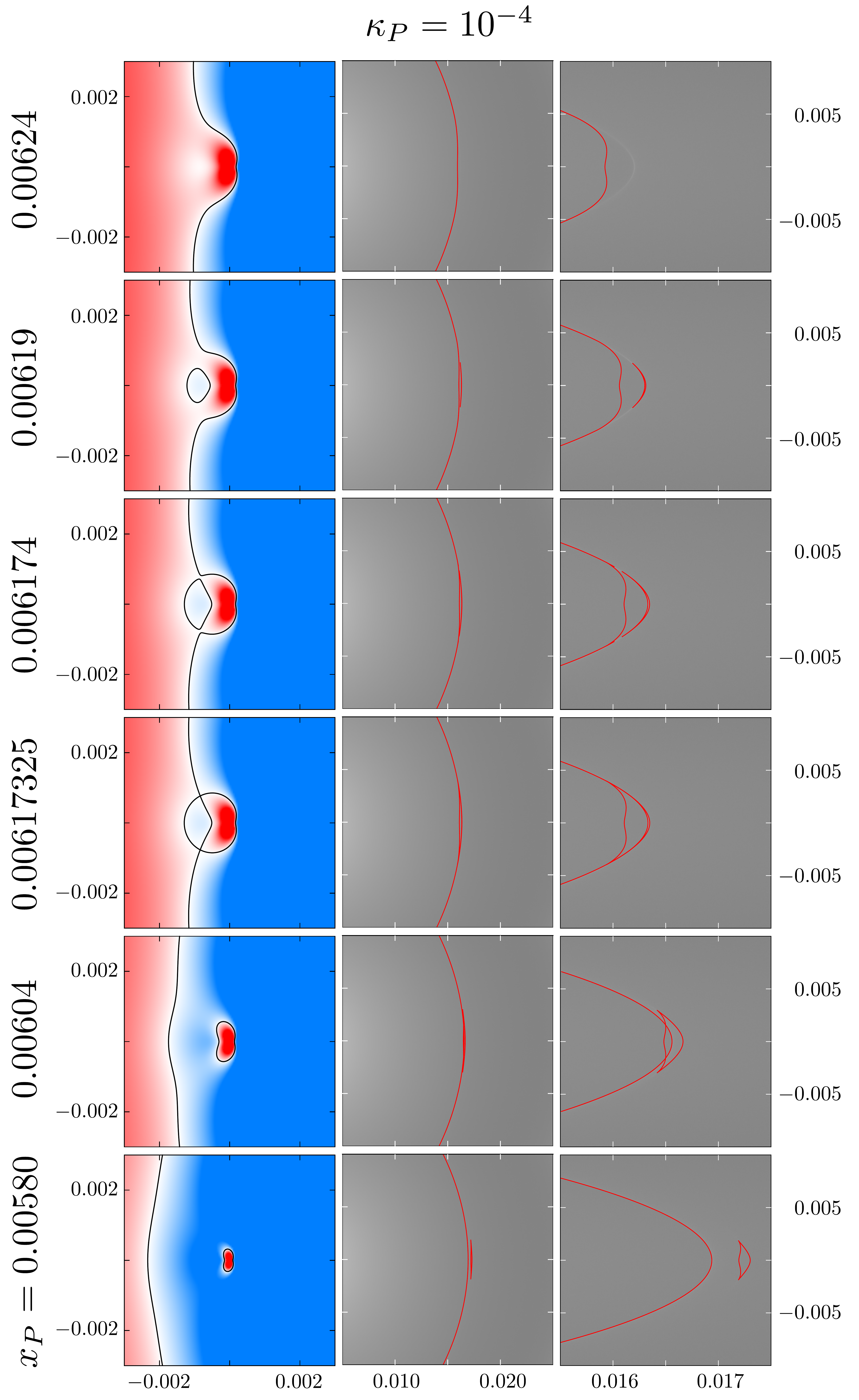}
\caption{Critical-curve and caustic details for a sub-critical $\kappa_{\text{P}}=10^{-4}$ point mass near the halo center. Shown for six radial positions $x_{\text{P}}$ marked along the left side, corresponding to transition across the adjacent dashed blue and orange boundaries along the left set of crosses in Figure~\ref{fig:transitions}. The critical-curve panels are fully inside the Einstein ring of the point mass. The caustic details shown in the central column are expanded in the right column $10\times$ horizontally to reveal the caustic structure. Notation and color bars as in Figure~\ref{fig:gallery-print}.}
\label{fig:origin_low-mass}}
\efi

\renewcommand{\thefigure}{8}
\bfi
{\centering
\vspace{0cm}
\hspace{0cm}
\includegraphics[width=8.5 cm]{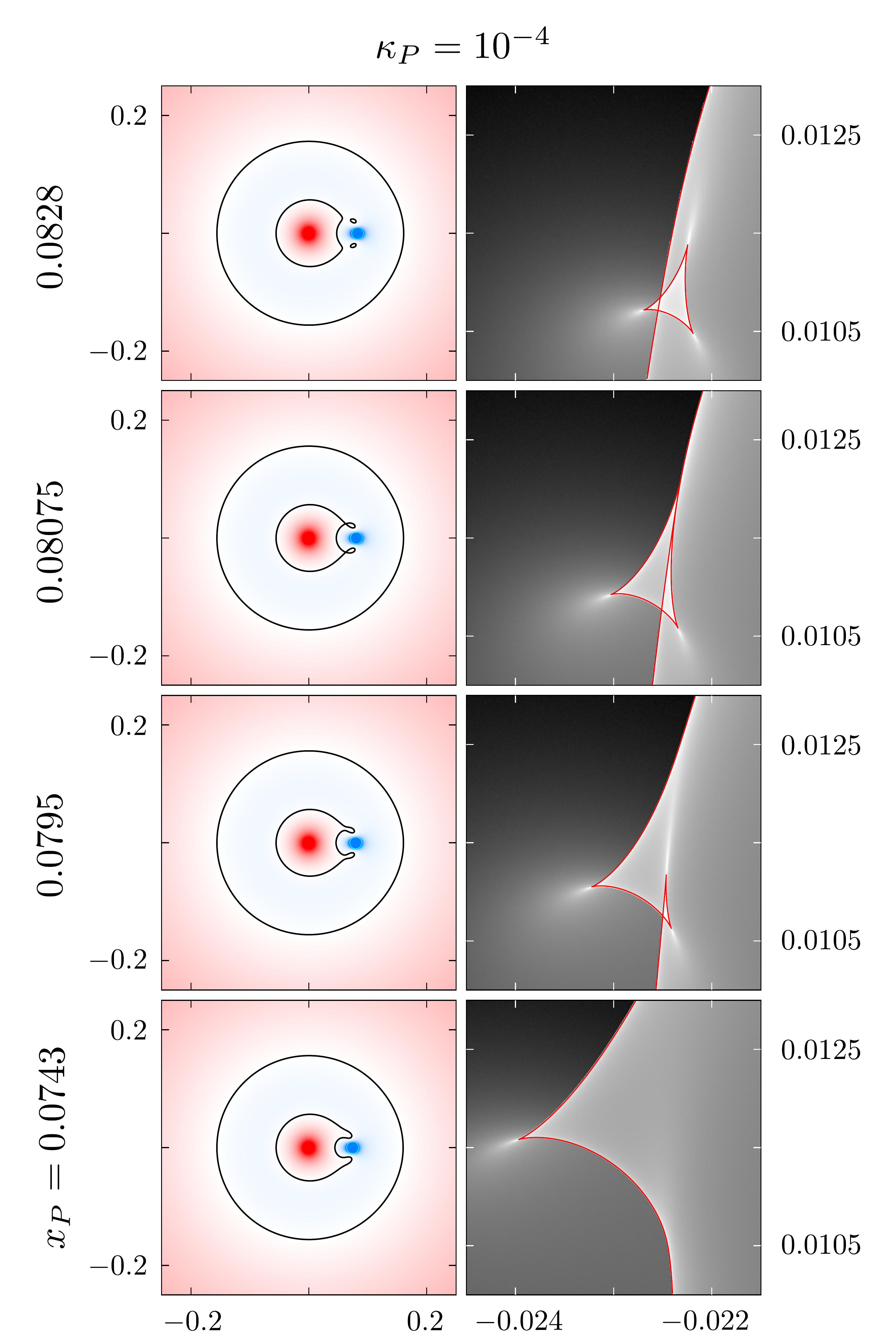}
\caption{Critical curves and caustic details for a $\kappa_{\text{P}}=10^{-4}$ point mass at two simultaneous beak-to-beak metamorphoses (third row from bottom). Shown for four radial positions $x_{\text{P}}$ marked along the left side, corresponding to transition across the lower solid blue boundary along the left set of crosses in Figure~\ref{fig:transitions}. Notation and color bars as in Figure~\ref{fig:gallery-print}.}
\label{fig:disconnecting_loops}}
\efi

After displacing the point mass away from the halo center, the first encountered boundaries are the adjacent dashed blue and orange lines corresponding to the beak-to-beak and lips metamorphoses, respectively. These are illustrated by the critical-curve and caustic details in Figure~\ref{fig:origin_low-mass}, with the entire sequence occurring between $x_{\text{P}}=0$ and $x_{\text{P}}=0.01$ in Figure~\ref{fig:gallery-online}. The additional third column shows the caustic detail from the second column with the horizontal scale expanded $10\times$ to reveal the caustic structure.

The critical-curve details from bottom to top reveal the small positive-Jacobian loop around the origin expanding and approaching the perturbed inner radial NFW critical curve, connecting with it symmetrically at two points in the third row for $x_{\text{P}}\approx0.00617325$, leaving a small negative-Jacobian loop, which shrinks and vanishes between the top two rows. The caustic plot in the bottom row shows the small two-cusped loop approaching the perturbed inner radial NFW caustic from outside. The two caustic loops overlap in the second row. In the third row two simultaneous beak-to-beak metamorphoses occur, in which the outer part of the two-cusped loop touches the perturbed inner radial NFW caustic at two points lying symmetrically above and below the horizontal axis. This pair-wise metamorphosis leads to a thin two-cusped crescent detaching from the caustic on the outer side. The small self-crossing features on the larger caustic in the fourth row vanish by the fifth row in two simultaneous swallow-tail metamorphoses. By the sixth row, the two-cusped crescent vanishes in a lips metamorphosis, leaving a barely noticeable higher-magnification trace in the magnification map.

The detachment of the two small critical-curve loops from the perturbed radial NFW critical curve that occurs between the second and the third row of Figure~\ref{fig:gallery-print} corresponds to a more complicated sequence of metamorphoses on the caustic, as shown in Figure~\ref{fig:disconnecting_loops}. The caustic detail presented in the right panels corresponds to the critical-curve detail below the point mass in the left panels. In the bottom row the caustic detail has a single cusp. By the second row it underwent a swallow-tail metamorphosis, which added a self-intersection and two cusps. A similar swallow-tail metamorphosis occurs on the caustic above the original cusp before the third row, adding a similar smaller feature. In the third row at $x_{\text{P}}\approx0.08075$ two simultaneous beak-to-beak metamorphoses occur, in which the facing cusps of the two swallow-tail features touch and reconnect. In the fourth row the caustic detail consists of a smooth fold and a detached three-cusped loop.

The transition following the detachment of the two loops also occurs between the second and the third row of Figure~\ref{fig:gallery-print}. The sequence shown in Figure~\ref{fig:elliptic_umbilic} corresponds to the green elliptic umbilic boundary in Figure~\ref{fig:transitions}. The two small critical-curve loops shrink to two points at $x_{\text{P}}\approx 0.095844$ in the second row, before expanding again to two loops. A similar effect can be seen in the caustic plots. The two small three-cusped loops shrink to two points at the elliptic umbilic metamorphosis in the second row, before expanding again to three-cusped loops.

\renewcommand{\thefigure}{9}
\bfi
{\centering
\vspace{0cm}
\hspace{0cm}
\includegraphics[width=8.5 cm]{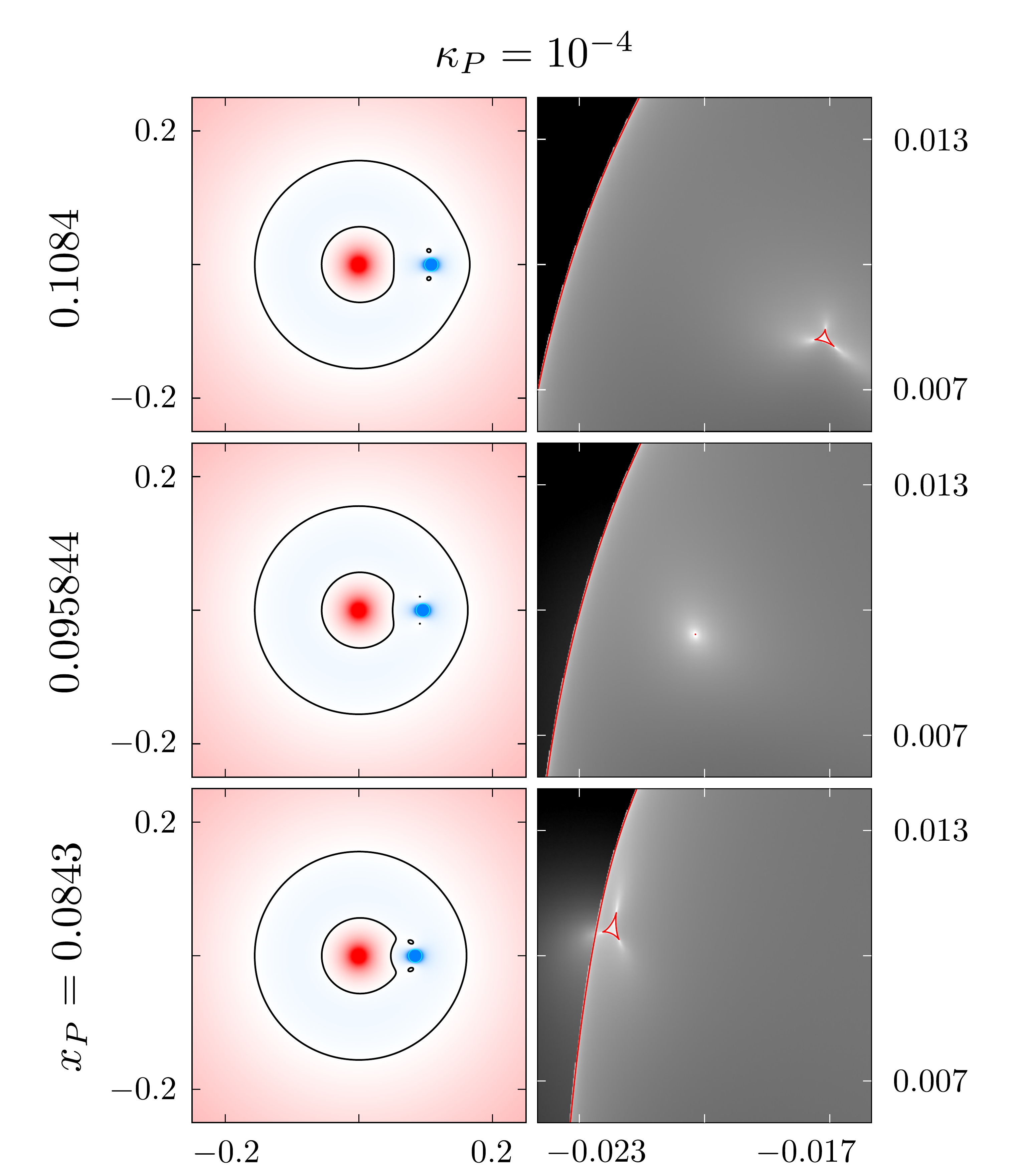}
\caption{Critical curves and caustic details for a $\kappa_{\text{P}}=10^{-4}$ point mass at two simultaneous elliptic umbilics (second row from bottom). Shown for three radial positions $x_{\text{P}}$ marked along the left side, corresponding to transition across the green boundary along the left set of crosses in Figure~\ref{fig:transitions}. Notation and color bars as in Figure~\ref{fig:gallery-print}.}
\label{fig:elliptic_umbilic}}
\efi

Along the right column of crosses in Figure~\ref{fig:transitions}, which corresponds to the super-critical $\kappa_{\text{P}}=10^{-3}$ example, the sequence is simpler, with the critical curve undergoing only five transitions as the point-mass position varies from the halo center to an asymptotic distance. All the transitions occurring from the second row to the top of Figure~\ref{fig:gallery-print} are the same as for the sub-critical example. The development from the first to the second row is much simpler than in the sub-critical case, with the small positive-Jacobian critical-curve loop around the origin directly expanding to form the perturbed NFW radial critical curve. The two-cusped caustic loop approaches the origin from the right, forming the crescent-like caustic seen in the first row, and extending further to form the perturbed NFW radial caustic.

\renewcommand{\thefigure}{10}
\bfi
{\centering
\vspace{0cm}
\hspace{0cm}
\includegraphics[width=8.5 cm]{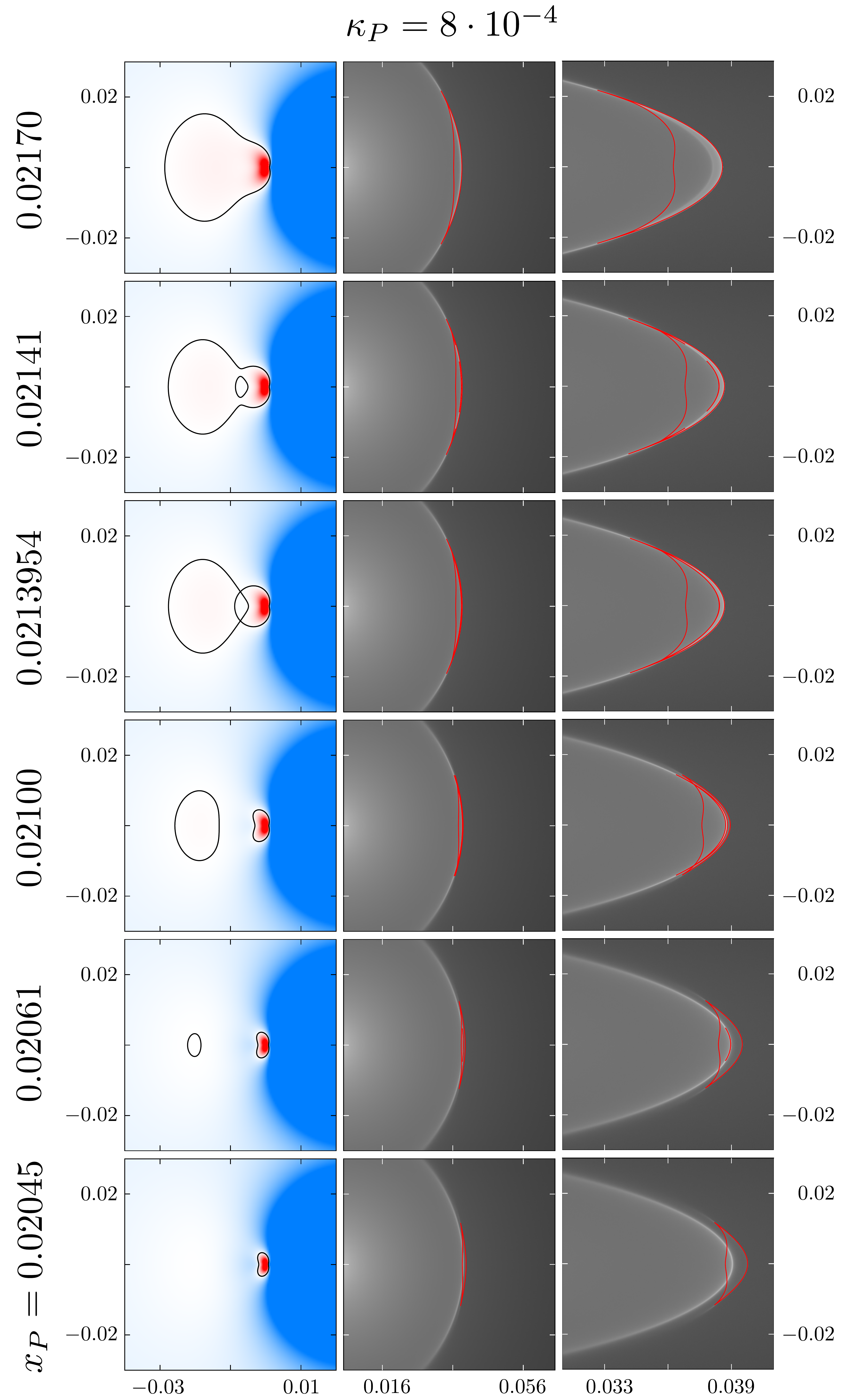}
\caption{Critical-curve and caustic details for a super-critical $\kappa_{\text{P}}=8 \cdot 10^{-4}$ point mass near the halo center. Shown for six radial positions $x_{\text{P}}$ marked along the left side, corresponding to transition across the orange, dashed blue, and orange boundaries along the $\kappa_{\text{P}}=8 \cdot 10^{-4}$ vertical line in the bottom right panel of Figure~\ref{fig:transitions}. The caustic details shown in the central column are expanded in the right column $6\times$ horizontally to reveal the caustic structure. Compare with the corresponding sub-critical sequence in Figure~\ref{fig:origin_low-mass}. Notation and color bars as in Figure~\ref{fig:gallery-print}.}
\label{fig:origin_high-mass}}
\efi

However, for super-critical mass parameters $\kappa_{\text{P}} \lesssim 9.86\cdot 10^{-4}$ the sequence of transitions close to the halo center differs markedly from either of the previous examples. For illustration, we demonstrate in Figure~\ref{fig:origin_high-mass} the transition across the orange, dashed blue, and orange boundaries along the $\kappa_{\text{P}}=8\cdot 10^{-4}$ grid line in the bottom right panel of Figure~\ref{fig:transitions}. These correspond to a sequence of lips, two simultaneous beak-to-beak, and lips metamorphoses. The additional third column in Figure~\ref{fig:origin_high-mass} shows the caustic detail from the second column with the horizontal scale expanded $6\times$ to reveal the caustic structure.

Between the first and the second row, a second small positive-Jacobian critical-curve loop appears to the left of the first one. The two loops expand and connect together symmetrically at two points in the fourth row for $x_{\text{P}}\approx0.0213954$. The detached inner small negative-Jacobian loop shrinks and vanishes between the top two rows. At larger $x_{\text{P}}$ the remaining positive-Jacobian loop forms the perturbed radial NFW critical curve. In the bottom row the caustic detail shows the small two-cusped loop that arrived from the right to the brighter ring in the magnification map. In the second row the caustic has a second thin two-cusped loop inside the first loop that appeared along the bright ring in a lips metamorphosis. The two caustic loops overlap on the third row. In the fourth row two beak-to-beak metamorphoses occur, in which the outer part of the first loop reconnects symmetrically at two points with the inner part of the second loop. In the fifth row the caustic displays a disconnected thin inner two-cusped loop and small self-crossing features on the outer caustic loop. The features vanish in simultaneous swallow-tail metamorphoses, followed by the vanishing of the small two-cusped loop in a lips metamorphosis. At larger $x_{\text{P}}$, the two-cusped caustic loop seen in the top row forms the perturbed radial NFW caustic.

\renewcommand{\thefigure}{11}
\bfi
{\centering
\vspace{-0.2cm}
\hspace{0cm}
\includegraphics[width=8.5 cm]{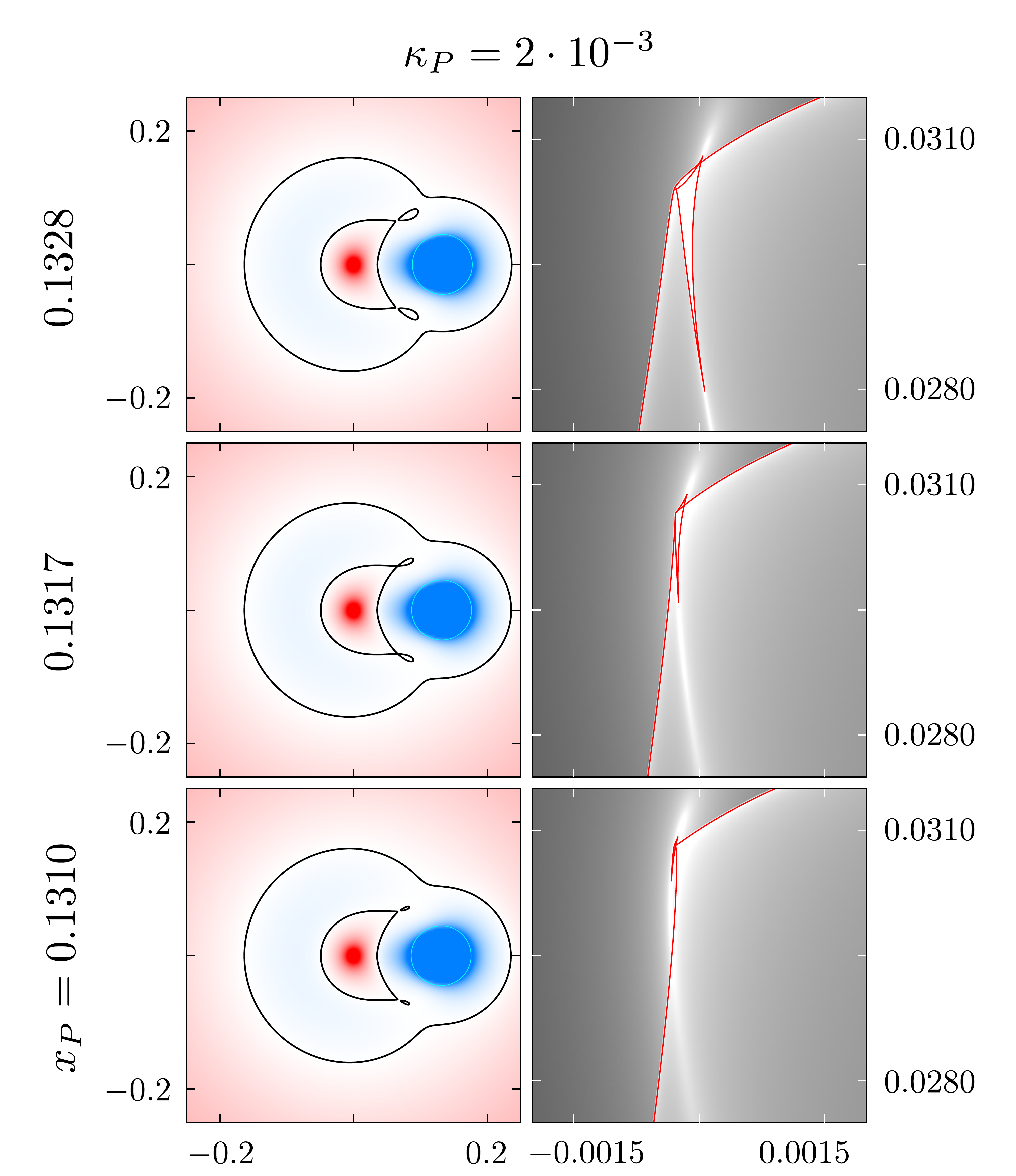}
\caption{Critical curves and caustic details for a $\kappa_{\text{P}}=2 \cdot 10^{-3}$ point mass at two simultaneous hyperbolic umbilics (second row from bottom). Shown for three radial positions $x_{\text{P}}$ marked along the left side, corresponding to transition across the violet boundary along the vertical $\kappa_{\text{P}}=2 \cdot 10^{-3}$ line in Figure~\ref{fig:transitions}. Notation and color bars as in Figure~\ref{fig:gallery-print}.}
\label{fig:hyperbolic_umbilic}}
\efi

For mass parameters $\kappa_{\text{P}} \gtrsim 1.323\cdot 10^{-3}$, higher than in the presented examples, the green boundary in Figure~\ref{fig:transitions} corresponding to the elliptic umbilic continues as the violet hyperbolic-umbilic boundary. We illustrate in Figure~\ref{fig:hyperbolic_umbilic} the transition across this boundary along the $\kappa_{\text{P}}=2\cdot 10^{-3}$ grid line in Figure~\ref{fig:transitions}. The two small critical-curve loops touch the perturbed NFW radial critical curve at $x_{\text{P}}\approx 0.1317$ in the second row, before detaching again, as seen on the third row. In the bottom row the caustic detail shows a single cusp on the larger caustic loop and a closely adjacent small two-cusped loop. At the hyperbolic-umbilic metamorphosis in the second row, both parts of the caustic touch at a blunt angular point. In this metamorphosis a cusp gets transferred from one part of the caustic to another. This can be seen in the third row, where the detached smaller loop has three cusps while there is no cusp on the larger perturbed NFW radial caustic.

For completeness, we mention the parabolic umbilic, which occurs for $\kappa_{\text{P}}\approx 1.323\cdot 10^{-3}$ at $x_{\text{P}}\approx0.1202$. At this point, which is marked by the black circle in Figure~\ref{fig:transitions}, boundaries corresponding to four different pair-wise metamorphoses meet, namely: elliptic umbilic, hyperbolic umbilic, lips, and beak-to-beak. When passing through this point in the parameter plot in the sense of increasing $x_{\text{P}}$, two small loops detach from cusp-like points on the perturbed radial NFW critical curve, starting as points and expanding like in the elliptic umbilic. During the underlying caustic metamorphosis the two cusps on the perturbed radial NFW caustic disappear as two small three-cusped caustic loops detach from them. These caustic loops also start as points and increase in size. Varying the point-mass parameters in the vicinity of the parabolic umbilic point leads to a variety of different changes to the local structure of the caustic \citep[e.g.,][]{godwin71,poston_stewart78}.

In Section~\ref{sec:central_position} we studied the structure of critical curves and caustics for centrally positioned point masses, and demonstrated the distinction between the sub- and super-critical cases. If we look at the transition sequences for different values of $\kappa_{\text{P}}$ in Figure~\ref{fig:transitions}, we see that the boundary intersections and limiting points form a finer subdivision into $\kappa_{\text{P}}$ intervals, each with a specific transition sequence. The approximate values separating these intervals are
\begin{multline}
\kappa_{\text{P}}\in\left\{0,\,1.62\cdot 10^{-4},\,2.71\cdot 10^{-4},\,9.86\cdot 10^{-4},\right.\\ \left. 1.32\cdot 10^{-3} ,\,2.26\cdot 10^{-3},\,2.97\cdot 10^{-3}\right\}\,.
\label{eq:kappa_P-intervals}
\end{multline}
To be precise, the first non-zero value actually should be replaced by two very close $\kappa_{\text{P}}$ values, because the cyan boundary in the bottom right panel of Figure~\ref{fig:transitions} intersects the orange before the adjacent dashed blue boundary. The eight different sequences of critical-curve transitions can be read off Figure~\ref{fig:transitions}, aided by the galleries in Figure~\ref{fig:gallery-print} and Figure~\ref{fig:gallery-online}, as well as the transition sequences described in this Section.

Note that the $\kappa_{\text{P}}=10^{-4}$ sub-critical sequence with all the described transitions lies in the first interval defined by the values in Equation~(\ref{eq:kappa_P-intervals}). Hence, any lower-mass point will have the same sequence of critical-curve transitions. Similarly, all point masses with $\kappa_{\text{P}} \gtrsim 2.97\cdot 10^{-3}$ share the same simple sequence, in which the only transition between the appearance of the small critical-curve loop around the origin and the final detachment of the point-mass loop from the perturbed tangential NFW critical curve is the hyperbolic umbilic. In this umbilic transition the perturbed radial and tangential NFW critical curves merely touch at two points. In the underlying caustic metamorphosis, the central four-cusped loop (similar to the caustic in the third row of Figure~\ref{fig:gallery-print}) extends and touches the two cusps of the crescent-like loop, resulting in a six-cusped central caustic and a smooth perturbed radial NFW caustic.

We point out that the interpretation of the parameter-space structure in Figure~\ref{fig:transitions} presented above focused on vertical transitions across the boundaries, which corresponds to placing a fixed-mass point at different positions in the halo. One may interpet the figure just as well by following transitions in the horizontal sense, corresponding to placing points of different masses at a fixed position in the halo. Such explorations reveal that the structure of the critical curve is more sensitive to the mass of the point when it is placed in certain regions of the halo, e.g., close to its center or in a part of the annular region between the radial and tangential NFW critical curves.

\section{DISCUSSION}
\label{sec:discussion}

Most of the presented results were computed for a single fiducial value of the halo convergence parameter $\kappa_{\text{s}}\approx 0.239035$, based on the \cite{merten_etal15} galaxy cluster data as shown in Figure~\ref{fig:clusters}. For these clusters, $\kappa_{\text{s}}$ varies from 0.15 to 0.84. Data from other cluster surveys such as OmegaWINGS \citep{biviano_etal17} indicate values as low as $\kappa_{\text{s}}\approx 0.01$. In addition, since $\kappa_{\text{s}}$ is normalized by the critical surface density defined by Equation~(\ref{eq:critical_density}), it also depends on the source redshift due to the direct proportionality $\kappa_{\text{s}} \propto D_{\text{ls}}/D_{\text{s}}$. Plots of this angular diameter distance ratio as a function of source redshift can be found in \cite{asada97} or \cite{umetsu20}.

The main question is what effect will changing $\kappa_{\text{s}}$ have on the parameter-space boundaries in Figure~\ref{fig:transitions}. While a systematic study is in progress, we mention here a few general properties. The overall pattern of the boundaries shrinks toward the origin of the parameter-space plot as $\kappa_{\text{s}}$ decreases, and expands from the origin as $\kappa_{\text{s}}$ increases. In the vertical direction this can be inferred from the scaling of the NFW critical-curve radii shown in Figure~\ref{fig:NFW_radii}, due to the importance of these radii on the vertical axis of Figure~\ref{fig:transitions}. In the horizontal direction this can be inferred from the importance of the relative ``mass ratio'' $\kappa_{\text{P}}/\kappa_{\text{s}}$ rather than the absolute value of $\kappa_{\text{P}}$. Nevertheless, while the change of the overall scale in the parameter-space plot is most prominent, the shapes of the boundaries change too, even though more weakly.

Regarding the source redshift, in addition to $\kappa_{\text{s}}$ it affects also the point-mass parameter $\kappa_{\text{P}}$. The critical surface density in the denominator of the last expression in Equation~(\ref{eq:kappa_P}) reveals that $\kappa_{\text{P}} \propto D_{\text{ls}}/D_{\text{s}}$, the same proportionality seen in $\kappa_{\text{s}}$. Indeed, it is the ratio $\kappa_{\text{P}}/\kappa_{\text{s}}=M_{\text{P}}/(\pi\,r_{\text{s}}^3\,\rho_{\text{s}})$ that is independent of the source, and is given purely by the properties of the lens. The third parameter of the studied lens system, $x_{\text{P}}$, does not depend on the source either.

It is worth pointing out the similarity of the structure of the perturbed tangential NFW critical curves and caustics (see Figure~\ref{fig:gallery-print}) to the critical curves and caustics of the two-point-mass lens \citep[e.g.,][]{erdl_schneider93,dominik99,pejcha_heyrovsky09}. For example, with increasing distance $x_{\text{P}}$ the caustic changes from a central four-cusped loop with two three-cusped loops, all of which merge to form a single six-cusped loop, which then splits into two four-cusped loops. The same sequence can be seen in the close, intermediate, and wide regimes of the two-point-mass lens. Even the two topmost beak-to-beak boundaries seen in Figure~\ref{fig:transitions} for $\kappa_{\text{P}}\lesssim 0.002$ resemble the two-point-mass parameter-space boundaries \citep[top left panel in their Figure 6]{erdl_schneider93}. In this mass range the lens systems differ in the ``fate'' of the three-cusped loops which recede from the central four-cusped loop as $x_{\text{P}}$ decreases. In the NFW + point-mass lens they get only as far as the perturbed radial NFW caustic and merge with it, while in the two-point-mass lens they escape to infinity as the separation of the points decreases to zero. Overall, the only caustic loop that escapes to infinity from both components of the NFW + point-mass lens is the small two-cusped loop receding as the point mass approaches the halo center, $x_{\text{P}}\to 0$.

The lens model explored in this work is relevant for several astrophysical scenarios. The primary motivation was to study the influence of a single galaxy on the overall lensing by a galaxy cluster. For this purpose, replacing the mass distribution within the galaxy by a point mass is the crudest possible approximation. Nevertheless, we may expect the structure of the critical curves to be similar except in the vicinity of the galaxy in the image plane. Clearly, the results in Figure~\ref{fig:transitions} will be more relevant for galaxies in a cluster that have lower relative masses and lower scale or cutoff radii of their mass distributions. In addition, we may expect different behavior for galaxies positioned close to the cluster center. Since the Jacobian of galactic mass distributions typically does not have a negative divergence, its combination with the positive Jacobian divergence of the cluster halo will yield different critical-curve structures than those demonstrated above for a point mass. A more complete understanding would be obtained by a comparison study using an extended mass distribution model for the galaxy.

The next astrophysical scenario for which our model is relevant is a (dwarf) satellite galaxy in the dark matter halo of a larger host galaxy. The comments made above for a galaxy within a cluster hold here too. For example, combining the host-galaxy halo with an extended-mass distribution for the satellite could lead to interesting comparisons with binary-galaxy lens models \citep[e.g.,][]{shin_evans08}. Our obtained results are more relevant for low-relative-mass compact satellite galaxies. In this scenario, another step toward a more realistic model would be to alter the spherical NFW model for the mass distribution of the host galaxy, e.g., by including ellipticity, adding a core radius, or altering the central density divergence \citep{evans_wilkinson98}.

Interestingly, our results are most relevant for the third scenario, a massive or super-massive black hole in the dark matter halo of its host galaxy. In such a setting the black hole is perfectly modeled by a simple point-mass lens, since the fraction of the lensed flux in relativistic higher-order images is negligible. In this scenario, the details close to the origin of the halo can be expected to change in mass-distribution models that eliminate the central divergence of the NFW density profile. Regarding the position of the black hole, in most cases it may be expected to lie at the center of the galactic halo. However, in dwarf galaxies massive black holes have been recently found even at their outskirts \citep{reines_etal20}. Similar wandering massive black holes are expected to inhabit the outer parts of regular galaxies as well \citep{guo_etal20}. Such cases give observational significance to the $x_{\text{P}}>0$ results. For the more typical central position, the main result is the difference between the lensing effects of sub- and super-critical-mass black holes.

\begin{equation*}
\quad
\end{equation*}

\section{SUMMARY}
\label{sec:summary}

We explored gravitational lensing by a massive object in a dark matter halo, using the simple model of a point mass in a halo with a NFW density profile. In this work we concentrated on the critical curves and caustics of the lens, in particular on their changes as a function of mass and position of the object. For computing the light deflection angle close to the NFW halo center we derived the numerically stable exact expression given by Equation~(\ref{eq:ln_f(x)_equivalent}).

For a point mass positioned at the center of the halo we demonstrated the existence of a critical mass parameter $\kappa_{\text{PC}}$, above which the gravitational influence of the object is strong enough to eliminate the radial critical curve and caustic of the NFW halo, as shown in Figure~\ref{fig:NFWP_radii}. This result is similar to the behavior of a point mass embedded in a cored isothermal sphere \citep{mao_etal01} and in a Plummer model \citep{werner_evans06}, but not in a singular isothermal sphere \citep{mao_witt12}. In Appendix~\ref{sec:Appendix-vanishing_curves}, we demonstrated the peculiar nature of the single radial critical curve and caustic in the critical-mass case, as well as the non-local metamorphosis of their vanishing. Crossing this caustic in the source plane does not change the number of images, and the magnification diverges on both sides of the caustic at a rate typical for a cusp approached perpendicularly, rather than for a fold caustic approached from its inner side.

For a general position of the point mass we found a surprising richness of critical-curve regimes in the point-mass parameter space. Transitions between the regimes occur along boundaries mapped in detail in Figure~\ref{fig:transitions}, which correspond to underlying caustic metamorphoses. The variety of local metamorphoses occurring in the low-mass regime of this simple lens model is unusually high. In fact, all caustic metamorphoses with three control parameters (beak-to-beak, swallow tail, lips, elliptic umbilic, and hyperbolic umbilic) plus the four-parameter parabolic umbilic occur here \citep{schneider_etal92}.

In Section~\ref{sec:discussion} we discussed the effect of changing the halo convergence parameter $\kappa_{\text{s}}$ on the presented results, and pointed out the similarities between the perturbed tangential NFW caustics for low masses and the caustic regimes of the two-point-mass lens. We commented on the relevance of the results for three different astrophysical scenarios: a galaxy in a galaxy cluster, a dwarf galaxy in the halo of a host galaxy, and a (super)massive black hole in a galaxy halo. Particularly in the first two cases, similar studies of lens models with an extended mass distribution for the smaller object can be performed for comparison with the presented results of the simplest model.


\begin{equation*}
\quad
\end{equation*}

We thank the anonymous referee and Paolo Salucci for helpful comments and suggestions on the manuscript. Work on this project was supported by Charles University Grant Agency project GA UK 1000218.


\onecolumngrid
\vskip 6mm
\twocolumngrid

\appendix

\section{ANALYTIC RESULTS AND APPROXIMATIONS}
\label{sec:Appendix-analytic}

Studies of gravitational lensing by NFW halos face analytic and numerical challenges due to the properties of the function $\mathcal{F}(x)$, defined here by Equation~(\ref{eq:f(x)}). The main difficulties occur close to the origin, for $x\ll 1$, where $\mathcal{F}(x)\to\infty$ but the combination $\ln(x/2)+\mathcal{F}(x)$, appearing in Equation~(\ref{eq:NFW_lens_equation}) and elsewhere, converges to zero. This problem becomes even more pronounced for lower values of the halo convergence parameter $\kappa_{\text{s}}$, when all the critical curves and caustics shrink exponentially fast to the origin, as shown in Equation~(\ref{eq:approx_x}), Equation~(\ref{eq:approx_y}), and Figure~\ref{fig:NFW_radii}. Even in double-precision arithmetic the expression from the first line of Equation~(\ref{eq:f(x)}) would fail to reproduce the results and transitions close to the origin presented in this work.

In the following Appendix~\ref{sec:Appendix-origin} we present expansions of different lensing quantities close to the origin to illustrate their local behavior. In addition, we present exact analytic expressions that do not suffer from the described cancellation problem. In Appendix~\ref{sec:Appendix-scale_radius} we illustrate the continuous and smooth nature of $\mathcal{F}(x)$ and the convergence $\kappa(x)$ across the scale radius, at $x=1$.

\subsection{Lensing near the origin}
\label{sec:Appendix-origin}

For $x\ll 1$ the four leading orders of the expansion of $\mathcal{F}(x)$ can be written as
\beq
\mathcal{F}(x)= -\ln{\frac{x}{2}}-\frac{x^2}{2}\,\ln{\frac{x}{2}}-\frac{x^2}{4} +\mathcal{O}(x^4\,\ln{x})\,,
\label{eq:f(x)_origin}
\eeq
which yields the two leading orders of the expression
\beq
\ln{\frac{x}{2}}+\mathcal{F}(x)= -\frac{x^2}{2}\,\ln{\frac{x}{2}}-\frac{x^2}{4} +\mathcal{O}(x^4\,\ln{x})\,,
\label{eq:ln_f(x)_origin}
\eeq
in which the logarithmic divergence of $\mathcal{F}(x)$ is cancelled and the combination shrinks to zero as $x^2\,\ln{x}$. By substituting Equation~(\ref{eq:f(x)_origin}) in Equation~(\ref{eq:NFW_kappa}) we get four leading orders of the NFW convergence expansion
\beq
\kappa(x)=-2\,\kappa_{\text{s}}\,(\,\ln{\frac{x}{2}}+1+\frac{3}{2}\,x^2\,\ln{x}+\frac{5}{4}\,x^2\,) +\mathcal{O}(x^4\,\ln{x})\,,
\label{eq:NFW_kappa_origin}
\eeq
showing its logarithmic divergence. The Jacobian behavior close to the origin can be obtained by substituting Equation~(\ref{eq:f(x)_origin}) in Equation~(\ref{eq:NFW_Jacobian}), yielding three leading orders of its expansion
\beq
\mathrm{det}\,J(\boldsymbol x)=4\,\kappa_{\text{s}}^2\,\ln^2{\frac{x}{2}} +4\,\kappa_{\text{s}}\,(1+2\,\kappa_{\text{s}})\,\ln{\frac{x}{2}}+1+4\,\kappa_{\text{s}}+3\,\kappa_{\text{s}}^2 +\mathcal{O}(x^2\,\ln{x})\,.
\label{eq:NFW_Jacobian_origin}
\eeq
Getting higher-order terms would require a higher-order expansion in Equation~(\ref{eq:f(x)_origin}). This result reveals the $\ln^2{x}$ divergence of the NFW Jacobian at the origin.

For computing the critical curves and caustics we need exact analytic expressions rather than series expansions. We first express the top row of Equation~(\ref{eq:f(x)}) in an equivalent form,
\beq
\mathcal{F}(x)=\frac{1}{\sqrt{1-x^2}}\,\ln{\frac{1+\sqrt{1-x^2}}{x}}\,,
\label{eq:f(x)_equivalent}
\eeq
which can then be combined with $\ln{(x/2)}$ to cancel the divergence at the origin,
\beq
\ln{\frac{x}{2}}+\mathcal{F}(x)=\frac{-x^2}{1-x^2+\sqrt{1-x^2}}\,\ln{\frac{x}{2}} +\frac{1}{\sqrt{1-x^2}}\,\ln{\frac{1+\sqrt{1-x^2}}{2}}\,.
\label{eq:ln_f(x)_equivalent}
\eeq
The first term reveals the leading order of the expansion at the origin as seen in Equation~(\ref{eq:ln_f(x)_origin}), while the second term contributes to higher orders starting from $\mathcal{O}(x^2)$. Both Equation~(\ref{eq:f(x)_equivalent}) and Equation~(\ref{eq:ln_f(x)_equivalent}) are valid for any $x<1$. These numerically stable expressions can be used in Equation~(\ref{eq:NFWP_Jacobian}) to compute the Jacobian and the critical curve, and in Equation~(\ref{eq:NFWP_lens_equation}) to compute the caustic.

\subsection{Lensing near the scale radius}
\label{sec:Appendix-scale_radius}

At the halo scale radius the $x<1$ expression in Equation~(\ref{eq:f(x)}) transitions smoothly to the $x>1$ expression. For illustration, we provide the three leading terms of the expansion valid on both sides of $x=1$,
\beq
\mathcal{F}(x)= 1-\frac{2}{3}\,(x-1)+\frac{7}{15}\,(x-1)^2+\mathcal{O}((x-1)^3)\,.
\label{eq:f(x)_radius}
\eeq
Using this result, we may expand the lens-equation combination
\beq
\ln{\frac{x}{2}}+\mathcal{F}(x)= 1-\ln{2}+\frac{x-1}{3}-\frac{(x-1)^2}{30}+\mathcal{O}((x-1)^3)\,.
\label{eq:ln_f(x)_radius}
\eeq
Substituting Equation~(\ref{eq:f(x)_radius}) in Equation~(\ref{eq:NFW_kappa}) yields two leading orders of the NFW convergence expansion
\beq
\kappa(x)=\kappa_{\text{s}}\,\left[\,\frac{2}{3}-\frac{4}{5}\,(x-1)\,\right]+\mathcal{O}((x-1)^2)\,.
\label{eq:NFW_kappa_radius}
\eeq
Getting higher-order terms would require a higher-order expansion in Equation~(\ref{eq:f(x)_radius}).

\section{VANISHING RADIAL CRITICAL CURVES AND CAUSTICS}
\label{sec:Appendix-vanishing_curves}

For a point mass with the critical value of the mass parameter $\kappa_{\text{P}}=\kappa_{\text{PC}}$ positioned centrally in a NFW halo, the single radial critical curve and single radial caustic have very unusual properties, as mentioned in Section~\ref{sec:central_position}. We illustrate the situation in Figure~\ref{fig:vanishing}, showing from top row to bottom the critical curve plotted over a color map of the Jacobian, the radial profile of the Jacobian, the caustic plotted over the total magnification map, and the radial profile of the total magnification in the vicinity of the (vanishing) radial caustics. The central column corresponds to the critical value $\kappa_{\text{P}}=\kappa_{\text{PC}}\approx 2.714 \cdot 10^{-4}$, which is bracketed in the left column by sub-critical $\kappa_{\text{P}} = 2 \cdot 10^{-4}$ and in the right column by super-critical $\kappa_{\text{P}} = 3 \cdot 10^{-4}$.

\renewcommand{\thefigure}{12}
\begin{figure*}[t]
{\centering
\vspace{0cm}
\hspace{0cm}
\includegraphics[height=17.5 cm]{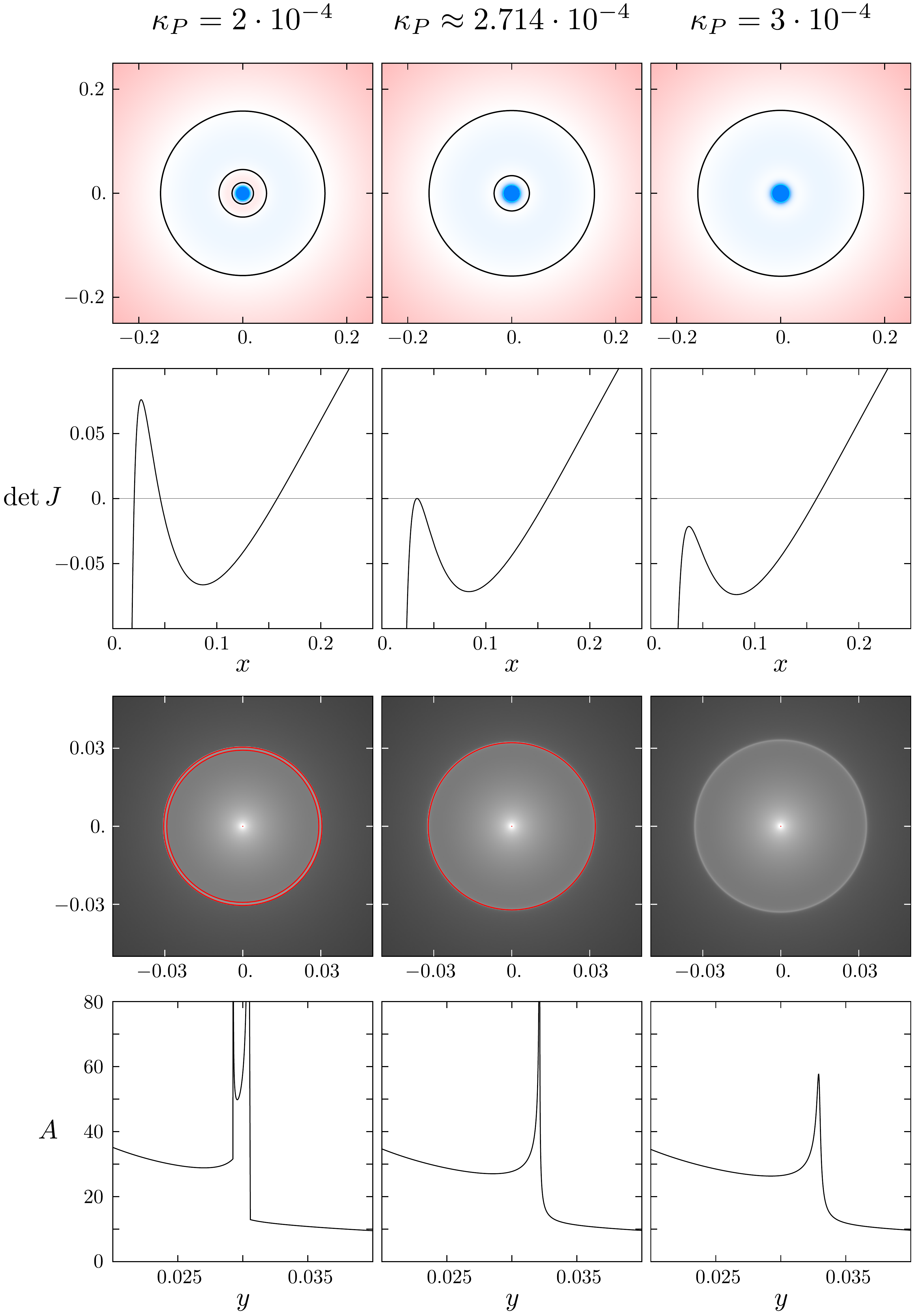}
\caption{Vanishing radial critical curves and caustics for a centrally positioned point mass. Rows from top: Jacobian maps with critical curves; radial profiles of Jacobian from halo center; total magnification maps with caustics; radial profiles of total magnification in vicinity of radial caustic position. Columns from left for increasing mass parameter: sub-critical $\kappa_{\text{P}} = 2 \cdot 10^{-4}$ (two radial critical curves and caustics); critical $\kappa_{\text{P}}=\kappa_{\text{PC}}\approx 2.714 \cdot 10^{-4}$ (single radial critical curve and caustic); super-critical $\kappa_{\text{P}} = 3 \cdot 10^{-4}$ (no radial critical curve and caustic). Note that the single radial caustic in the central column is not a fold caustic (see Appendix~\ref{sec:Appendix-vanishing_curves}). Notation and color bars as in Figure~\ref{fig:gallery-print}. }
\label{fig:vanishing}}
\end{figure*}

Along the critical curve the Jacobian is zero by definition, but for $\kappa_{\text{P}}=\kappa_{\text{PC}}$ it is negative on both sides of the radial critical curve, as seen in the top two panels of the central column. Thus, images on both sides have the same (negative) parity. The caustic, which is shown in the third panel of the central column, is even more peculiar. Along the caustic the magnification is infinite by definition. However, the number of images of a point source does not change when crossing this radial caustic. What's more, the magnification is divergent from both sides of the radial caustic, as seen in the bottom panel of the central column. This is unlike the usual fold caustic, across which the magnification changes discontinuously, diverging when approached from the inner side but reaching a finite value when approached from the outer side.

We are not aware of any previous example in the gravitational lensing literature of a smooth caustic curve that is not a fold caustic. The same type of caustic clearly should appear even for a critical value of the central mass embedded in a cored isothermal \citep{mao_etal01} or a Plummer \citep{werner_evans06} density profile. In addition, a similar effect occurs in certain non-gravitational plasma lens models \citep{er_rogers18}. However, to our knowledge the peculiar nature of such a caustic has not been described yet.

The vanishing of the radial critical curves and caustics presents a unique type of caustic metamorphosis. The usual metamorphoses, such as those discussed in Section~\ref{sec:transition_details}, occur at a single point. Their properties are studied by Taylor-expanding the lens potential and lens equation in the vicinity of the point. This metamorphosis is not point-like; it occurs along the full length of the caustic simultaneously, as a consequence of the axial symmetry of the lens configuration. Nevertheless, it is related to the common beak-to-beak metamorphosis, in which two facing fold caustics approach each other, touch, and reconnect, forming two facing cusps that recede from the metamorphosis point. Here the two facing radial caustics are perfectly parallel, hence they come into contact and the metamorphosis occurs simultaneously along their full length. Instead of forming receding cusps, which do not arise here due to the symmetry, the caustics simply vanish, leaving a ring-like maximum in the magnification map as well as a ring-like maximum in the negative-Jacobian surroundings in the image plane.

This interpretation is supported by the character of the total-magnification divergence at the caustic. For the critical value of $\kappa_{\text{P}}$ we find $A(y)\sim|\,y-y_{\text{PR}}\,|^{-2/3}$ on both sides of the caustic radius $y_{\text{PR}}$, as shown in the bottom panel of the central column. This is the generic magnification decline perpendicular to the axis of a cusp, as seen for example by setting $y_\parallel=0$ in equation~(A6) of \cite{pejcha_heyrovsky09}. For $\kappa_{\text{P}} = 2 \cdot 10^{-4}$ the divergence at the outer radial caustic follows $A(y)\sim(y_{\text{PR1}}-y)^{-1/2}$ for $y<y_{\text{PR1}}$, and the divergence at the inner radial caustic follows $A(y)\sim(y-y_{\text{PR2}})^{-1/2}$ for $y>y_{\text{PR2}}$, as shown in the bottom panel of the left column. Both are generic fold caustics oriented inside the annulus enclosed by them. Note that the inner radial caustic is weaker than the outer one, as indicated by the more narrow divergence.

The critical value of the mass parameter $\kappa_{\text{PC}}\approx 2.714 \cdot 10^{-4}$ corresponds to the fiducial NFW halo convergence parameter $\kappa_{\text{s}}\approx 0.239035$ used in this work. However, the occurrence of such a critical mass and the accompanying metamorphosis is generic, with the critical value depending on the halo in which the mass is embedded: $\kappa_{\text{PC}}=\kappa_{\text{PC}}(\kappa_{\text{s}})$.

\end{document}